\pdfoutput=1
\newcommand*{\ATLASLATEXPATH}{}
\documentclass[PAPER, cernpreprint,texlive=2017,english]{\ATLASLATEXPATH atlasdoc}
 
\usepackage[backend=biber,biblatex=true]{\ATLASLATEXPATH atlaspackage}
\usepackage{pdflscape}
\usepackage{amssymb}
\usepackage{pifont}
\usepackage{longtable}
\usepackage{hyperref}
\usepackage{rotating}
\usepackage[backref=false]{\ATLASLATEXPATH atlasbiblatex}
 
\usepackage{\ATLASLATEXPATH atlasphysics}
 
\graphicspath{{logos/}{figures/}}
 
\usepackage{ANA-SUSY-2018-31-PAPER-defs}

\AtlasTitle{Search for bottom-squark pair production with the ATLAS detector in final states containing Higgs bosons, $b$-jets and missing transverse momentum}

\AtlasAbstract{
The result of a search for the pair production of the lightest supersymmetric partner of the bottom quark (\sbot) using $139\,\ifb$ of proton--proton data collected at $\sqrt{s} = 13$\,TeV by the ATLAS detector is reported. In the supersymmetric scenarios considered both of the bottom-squarks decay into a $b$-quark and the second-lightest neutralino, $\sbot \rightarrow b + \neuttwo$. Each \neuttwo is assumed to subsequently decay with 100\% branching ratio into a Higgs boson ($h$) like the one in the Standard Model and the lightest neutralino: $\neuttwo \rightarrow h + \neut$. The $\neut$ is assumed to be the lightest supersymmetric particle (LSP) and is stable. Two signal mass configurations are targeted: the first has a constant LSP mass of 60\,GeV; and the second has a constant mass difference between the $\neuttwo$ and $\neut$ of 130\,GeV. The final states considered contain no charged leptons, three or more \bjets, and large missing transverse momentum. No significant excess of events over the Standard Model background expectation is observed in any of the signal regions considered. Limits at the 95\% confidence level are placed in the supersymmetric models considered, and bottom-squarks with mass up to 1.5\,TeV are excluded. }
 
\author{The ATLAS Collaboration}

\AtlasRefCode{SUSY-2018-031}
 
\PreprintIdNumber{CERN-EP-2019-142}
 
\AtlasDate{\today}

\journal{JHEP}
\AtlasJournalRef{\JHEP 12 (2019) 060}
\AtlasDOI{10.1007/JHEP12(2019)060}

\AtlasCoverSupportingNote{ANA-SUSY-2018-31-INT}{https://cds.cern.ch/record/2638536}
 
\AtlasCoverCommentsDeadline{9 July 2019}
 
\AtlasCoverAnalysisTeam{John Anders, Davide Costanzo, Monica D'Onofrio, Evangelos Kourlitis, Calum Macdonald, Hamish Teagle, Thomas Weston}
 
\AtlasCoverEdBoardMember{Stan Lai (chair), Rui Wang, Stefano Passagio}

\AtlasCoverEgroupEditors{atlas-ana-susy-2018-31-analysis-team@cern.ch}
 
\AtlasCoverEgroupEdBoard{atlas-ana-susy-2018-31-editorial-board@cern.ch}
 
\hypersetup{pdftitle={ATLAS document},pdfauthor={The ATLAS Collaboration}}
 
\begin{document}
 
\maketitle

\section{Introduction}
\label{sec:intro}
Supersymmetry (SUSY)~\cite{Golfand:1971iw,Volkov:1973ix,Wess:1974tw,Wess:1974jb,Ferrara:1974pu,Salam:1974ig} provides an extension
to the Standard Model (SM) that solves the hierarchy problem~\cite{Sakai:1981gr,Dimopoulos:1981yj,Ibanez:1981yh,Dimopoulos:1981zb}
by introducing partners of  the known bosons and fermions. In $R$-parity-conserving models~\cite{Farrar:1978xj},
SUSY particles are produced in pairs and the lightest supersymmetric particle (LSP) is stable and provides a candidate
for dark matter~\cite{Goldberg:1983nd,Ellis:1983ew}. The superpartners of the SM bosons (the wino, bino and higgsinos) mix to form the neutralinos ($\tilde{\chi}_{1,2,3,4}^{0}$) and charginos ($\tilde{\chi}_{1,2}^{\pm}$) physical states.
For a large selection of models, the LSP is the lightest neutralino (\neut).
Naturalness considerations suggest that the supersymmetric partners of the third-generation quarks are light~\cite{Barbieri:1987fn,deCarlos:1993yy}. If this is assumed, the lightest bottom-squark ($\bone$) and lightest top-squark ($\tone$) mass eigenstates\footnote{The scalar partners of the left-handed and right-handed chiral components of the bottom
quark ($\tilde{b}_{\mathrm{L},\mathrm{R}}$) or top quark ($\tilde{t}_{\mathrm{L},\mathrm{R}}$) mix to form mass eigenstates
for which \bone and \tone are defined as the lighter of the two states.} could be significantly lighter than the other squarks
and the gluinos.
As a consequence, $\bone$ and $\tone$ could be pair-produced with relatively large cross-sections at the Large Hadron Collider (LHC). Depending on the mass hierarchy considered, it is possible that the \bone\ and \tone\ could decay into final states with Higgs bosons, \emph{h}, like the one in the SM, and this allows the Higgs boson to be used as a probe for new physics.
 
This article presents a search for the pair production of bottom squarks decaying into the LSP via a complex decay chain containing the second-lightest neutralino (\neuttwo) and the Higgs boson: $\sbot \rightarrow b + \neuttwo$ and subsequently $\neuttwo \rightarrow h + \neut$. Such a decay hierarchy is predicted in minimal supersymmetric extensions to the SM (MSSM)~\cite{Fayet:1976et,Fayet:1977yc}, with $h$ assumed to be the lightest of the neutral bosons introduced in the MSSM.
The bottom squark decaying through a next-to-lightest neutralino is one of the possible modes within the MSSM. Dedicated searches for direct decays into the lightest neutralino ($\sbot \rightarrow b\neut$) or a chargino ($\sbot \rightarrow t \tilde{\chi}_{1}^{\pm}$) have been reported by the ATLAS and CMS collaborations (see for example ~\cite{SUSY-2016-28,CMS-SUS-16-036} and ~\cite{SUSY-2016-14, CMS-SUS-16-035,CMS-SUS-16-041}).
 
When the LSP is bino-like and the \neuttwo is a wino--higgsino mixture, the branching ratio ($\mathcal{B}$) of $\neuttwo \rightarrow h + \neut$ is enhanced relative to the other possible \neuttwo decays. The Higgs boson mass is taken to be 125~\GeV, and the decay into a pair of bottom quarks is assumed to be the same as in the SM ($\mathcal{B}$ = 58\%~\cite{ATLAS-CONF-2018-036,deFlorian:2016spz}), although it could be enhanced or reduced in the MSSM.
 
\begin{figure}[!htb]
\centering
\includegraphics[width=0.45\columnwidth]{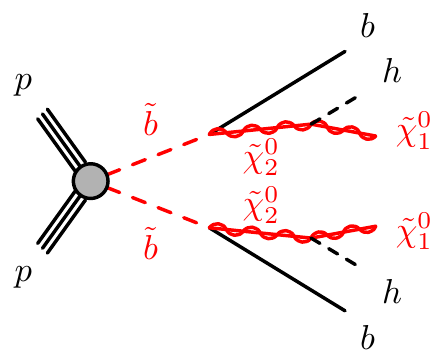}
\caption{Graphical representation of the SUSY signal targeted by this analysis. Bottom squarks are produced in pairs
and subsequently decay into $b\neuttwo$ with $\mathcal{B}$ = 100\%. The two $\neuttwo$ particles decay into $h\neut$ also with $\mathcal{B}$ = 100\%. }
\label{fig:sbsb-bbhhN1N1}
\end{figure}

This search is interpreted within simplified model scenarios~\cite{Alwall:2008ve, Alwall:2008ag} and Figure~\ref{fig:sbsb-bbhhN1N1} illustrates the targeted model. In the first set of models, already considered by the ATLAS Collaboration using 8~TeV data \cite{Aad:2015pfx}, the mass of the \neut is fixed at 60~\GeV. The bottom-squark and \neuttwo masses vary in the ranges 250--1600~\GeV\ and 200--1500~\GeV, respectively. The assumption about the \neut mass is motivated by dark-matter relic density measurements and might be favoured in Higgs-pole annihilation scenarios~\cite{Escudero:2016gzx} where $m_{\neut} \simeq m_h/2 $. The previous search performed by ATLAS using 8~TeV data excluded bottom-squark masses up to 750~\GeV\ in this scenario~\cite{Aad:2015pfx}.
 
The second set of SUSY models assumes a fixed mass difference between the \neuttwo and \neut, sufficient to produce an on-shell Higgs boson. The mass difference, \dmNN, is set to 130~\GeV, whilst bottom-squark and \neut masses vary in the ranges 400--1500~\GeV\ and 1--800~\GeV, respectively. A similar scenario is considered by the CMS Collaboration in Ref.~\cite{Sirunyan:2019iwo}, 
where the $h \rightarrow \gamma\gamma$ decay mode is exploited to exclude bottom-squark masses up to 530~\GeV; 
no prior ATLAS searches have targeted these models.
 
The final states are characterised by a unique signature, which contains many jets, of which up to six can be identified as originating from the fragmentation of \bquarks (referred to as \bjets), missing transverse momentum ($\ptmiss$, the magnitude thereof referred to as \etmiss), and no charged leptons (referred to as leptons). New selections and dedicated procedures aiming to maximise the efficiency of reconstructing the Higgs boson candidates decaying into a \bquark\ pair are employed in this article. Section~\ref{sec:detector} presents a brief overview of the ATLAS detector, with Section~\ref{sec:data_and_samples} describing the data and simulated samples used in the analysis. The event reconstruction methods are explained in Section~\ref{sec:reconstruction}. An overview of the analysis strategy is presented in Section~\ref{sec:event_selection}, with the background estimation strategy discussed in Section~\ref{sec:background_estimation}. The systematic uncertainties considered in the analysis are described in Section~\ref{sec:systematic_uncertainties}. Section~\ref{sec:results_and_interpretation} presents the results and interpretation thereof, with the conclusions presented in Section~\ref{sec:conclusion}.
 
 
\section{ATLAS detector}
\label{sec:detector}
The ATLAS detector~\cite{PERF-2007-01} is a multipurpose particle physics detector with a forward--backward symmetric cylindrical
geometry and nearly 4$\pi$ coverage in solid angle.\footnote{ATLAS uses a right-handed coordinate system with its origin at the nominal
interaction point in the centre of the detector. The positive $x$-axis is defined by the direction from the interaction point to the centre
of the LHC ring, with the positive $y$-axis pointing upwards, while the beam direction defines the $z$-axis. Cylindrical coordinates
$(r,\phi)$ are used in the transverse plane, $\phi$ being the azimuthal angle around the $z$-axis. The component of momentum in the transverse plane is denoted by \pt. The pseudorapidity $\eta$ is
defined in terms of the polar angle $\theta$ by $\eta=-\ln\tan(\theta/2)$. Rapidity is defined as $y$ = $0.5 \ln[(E + p_z)/(E-p_z)]$ where $E$ denotes the energy, and $p_z$ is the component of the momentum along the beam direction. The separation of two objects in $\eta$--$\phi$ space is given by $\Delta R=\sqrt{(\Delta \eta)^2 + (\Delta \phi)^2}$.}
The inner tracking detector consists of pixel and silicon microstrip detectors
covering the pseudorapidity region $|\eta| < 2.5$, surrounded by a transition radiation tracker
which enhances electron identification in the region $|\eta| < 2.0$.
Between Run 1 and Run 2, a new inner pixel layer, the insertable B-layer~\cite{IBL,PIX-2018-001}, was added at a mean sensor radius of
3.3~cm. The inner detector is surrounded by a thin superconducting solenoid providing an axial 2~T magnetic field and by a fine-granularity lead/liquid-argon (LAr) electromagnetic calorimeter covering $|\eta| < 3.2$. A steel/scintillator-tile calorimeter provides hadronic coverage in the central pseudorapidity range ($|\eta| < 1.7$).
The endcap and forward regions ($1.5 < |\eta| < 4.9$) of the hadronic calorimeter are made of LAr active layers with either copper or tungsten
as the absorber material.
An extensive muon spectrometer with an air-core toroidal magnet system surrounds the calorimeters.
Three layers of high-precision tracking chambers provide coverage in the range $|\eta| < 2.7$, while dedicated fast chambers allow triggering in the region $|\eta| < 2.4$. The ATLAS trigger system consists of a hardware-based level-1 trigger followed by a software-based high-level trigger~\cite{TRIG-2016-01}.

\section{Data and simulated event samples}
\label{sec:data_and_samples}
The data analysed in this study correspond to a total of $139\,\ifb$ of proton--proton ($pp$) collision data collected by the ATLAS detector with a centre-of-mass energy of 13~\TeV\ and a 25\,ns proton bunch crossing interval in the period between 2015 and 2018. All detector subsystems were required to be operational during data taking.
The average number of interactions per bunch crossing (pile-up) increased from \averagemu$=20$ (2015--2016 dataset)
to \averagemu$=37$ (2018 dataset), with a highest \averagemu$=38$ (2017 dataset).
The uncertainty in the combined 2015--2018 integrated luminosity is 1.7\,\% \cite{ATLAS-CONF-2019-021}, obtained using the LUCID-2 detector \cite{LUCID2} for the primary luminosity measurements.

Events are required to pass an $\etmiss$\ trigger~\cite{ATL-DAQ-PUB-2016-001} which is fully efficient for events with reconstructed $\etmiss > 250~\GeV$. Additional single-lepton triggers requiring electrons or muons are used to estimate the SM backgrounds, with an offline selection of $\pt(\ell) > 27$~\GeV\ used to ensure the trigger is fully efficient ($\ell = e, \mu$).
 
Dedicated Monte Carlo (MC) simulated samples are used to model SM processes and estimate the expected signal yields.
All samples were produced using the ATLAS simulation infrastructure~\cite{SOFT-2010-01} and GEANT4~\cite{Agostinelli:2002hh},
or a faster simulation based on a parameterisation of the calorimeter response and GEANT4 for the other detector
systems~\cite{SOFT-2010-01}. 
 
The SUSY signal samples were generated with \madgraph~v2.6.2~\cite{Alwall:2014hca} at leading order (LO) and
interfaced to \pythia~v8.230~\cite{Sjostrand:2014zea} for the modelling of the parton showering (PS), hadronisation
and the underlying event with the A14~\cite{ATL-PHYS-PUB-2014-021} set of tuned parameters (tune).
The matrix element (ME) calculation was performed at tree level and includes the emission of up to two additional partons.
The ME--PS matching was done using the CKKW-L~\cite{Lonnblad:2012ix} prescription, with a matching scale set
to one quarter of the bottom-squark mass.
The NNPDF2.3 LO~\cite{Ball:2012cx} parton distribution function (PDF) set was used. Signal cross-sections were calculated to approximate next-to-next-to-leading order in the strong coupling constant, adding the resummation of soft gluon emission at next-to-next-to-leading-logarithm (approximate NNLO+NNLL)~\cite{Beenakker:2016lwe,Beenakker:1997ut,Beenakker:2010nq,Beenakker:2016gmf} accuracy. The nominal cross-section and its uncertainty were derived using the PDF4LHC15\_mc PDF set, following the recommendations of Ref.~\cite{Butterworth:2015oua}. For $\bone$ masses between 400~GeV and 1.5~TeV, the cross-sections range from 2.1\,pb to 0.26\,fb, with uncertainties from 6\% to 17\%.

The SM backgrounds considered in this analysis are: \ttbar pair production; single-top-quark production; \Zjets; \Wjets; \ttbar production with an electroweak ($ttV$) or Higgs ($ttH$) boson; and diboson production. The samples were simulated using different MC generator programs depending on the process.
Pair production of top quarks, $t\bar{t}$, was generated using \powhegbox~v2~\cite{Frixione:2007nw,Nason:2004rx,Frixione:2007vw,Alioli:2010xd} 
interfaced with \pythia~v8.230 and the A14 tune with the NNPDF2.3 LO PDF set for the ME calculations.
The $h_\mathrm{damp}$ parameter in \powhegbox, which controls the \pt\ of the first additional emission
beyond the Born level and thus regulates the \pt\ of the recoil emission against the $t\bar{t}$ system,
was set to 1.5 times the top-quark mass ($m_t~=~172.5$~\GeV) as a result of studies documented in Ref.~\cite{ATL-PHYS-PUB-2016-016}.
The generation of single top quarks in the $Wt$-channel, $s$-channel and $t$-channel production modes was performed by
\powhegbox~v2~\cite{Re:2010bp,Nason:2004rx,Frixione:2007vw,Alioli:2010xd} similarly to the $t\bar{t}$ samples.  For all processes involving top quarks, top-quark spin
correlations were preserved. All events with at least one leptonically decaying $W$ boson were retained;
fully hadronic $\ttbar$ and single-top events do not contain sufficient \etmiss\ to contribute significantly
to the background.
The production of $\ttbar$ pairs in association with electroweak vector bosons ($W, Z$) or Higgs bosons
was modelled by samples generated at NLO using \madgraph~v2.2.3 and showered with \pythia~v8.212.
Events containing $W$ or $Z$ bosons with associated jets, including jets from the fragmentation of
heavy-flavour quarks, were simulated using the \sherpa~v2.2.1~\cite{Gleisberg:2008ta} generator.
Matrix elements were calculated for up to two additional partons at NLO and four partons
at LO using the \comix~\cite{Gleisberg:2008fv} and \openloops~\cite{Cascioli:2011va} ME
generators and were merged with the \sherpa PS~\cite{Schumann:2007mg} using the ME+PS@NLO prescription~\cite{Hoeche:2012yf}.
The NNPDF3.0 NNLO~\cite{Ball:2012cx} PDF set was used in conjunction with a dedicated PS tune developed by the \sherpa authors.
Diboson processes were also simulated using the \sherpa generator using the NNPDF3.0 NNLO PDF set.
They were calculated for up to one ($ZZ$) or zero ($WW, WZ$) additional partons at NLO and up to three additional partons at LO.
Other potential sources of backgrounds,
such as the production of three or four top quarks or three gauge bosons, are found to be negligible.
Finally, contributions from multijet background are estimated from data using a jet smearing procedure described in Ref.~\cite{jetsmear}
and are found to be negligible in all regions.
 
All background processes are normalised to the best available theoretical calculation for their
respective cross-sections. The NLO \ttbar\ inclusive production cross-section is corrected to the theory
prediction at NNLO in QCD including the resummation of NNLL soft-gluon terms calculated using
\textsc{Top++2.0}~\cite{Beneke:2011mq,Cacciari:2011hy,Baernreuther:2012ws,Czakon:2012zr,Czakon:2012pz,Czakon:2013goa,Czakon:2011xx}.
Samples of single-top events are normalised to the NLO cross-sections reported
in Refs.\,\cite{Kidonakis:2011wy,Kidonakis:2010tc,Kidonakis:2010ux}.
 
For all samples, except those generated using \sherpa, the \evtgen~v1.2.0~\cite{Lange:2001uf} program was used to simulate the properties of
the bottom- and charm-hadron decays.
All simulated events include a modelling of contributions from pile-up by overlaying minimum-bias $pp$ interactions from the same (in-time pile-up) and
nearby (out-of-time pile-up) bunch crossings simulated in \pythia~v8.186 and \evtgen~v1.2.0 with the A3~\cite{ATL-PHYS-PUB-2016-017}
tune and the NNPDF2.3 LO set~\cite{Ball:2012cx}.

 
\section{Event reconstruction}
\label{sec:reconstruction}
This search is based upon a selection of events with many \bjets, large missing transverse momentum and no charged leptons (electrons and muons) in the final state. 
All events are required to have a reconstructed primary vertex which is consistent with the beamspot envelope and consists of at least two associated tracks in the inner detector with \pt$ > $~500\,MeV. If more than one vertex passing the above requirements is found, the one with the largest sum of the squares of transverse momenta of associated tracks~\cite{ATL-PHYS-PUB-2015-026} is chosen.
 
Jet candidates are reconstructed from three-dimensional clusters of energy in the calorimeter~\cite{PERF-2014-07} with the anti-$k_t$ jet algorithm~\cite{Cacciari:2008gp, Cacciari:2011ma} using a radius parameter of 0.4. The application of a jet energy scale (JES) correction derived from data and simulation~\cite{PERF-2016-04} is used to calibrate the reconstructed jets. A set of quality criteria is applied to identify jets which arise from non-collision sources or detector noise~\cite{ATLAS-CONF-2015-029} and any event which contains a jet failing to satisfy these criteria is removed. Additional jets that arise from pile-up interactions are rejected by applying additional track-based selections to jets with $\pt < 120$~\GeV\ and $|\eta| < 2.4$~\cite{ATLAS-CONF-2014-018}, and the jet momentum is corrected by subtracting the expected average energy contribution from pile-up using the jet area method~\cite{ATLAS-CONF-2017-065}. Jets are classified as either `baseline' or `signal'; baseline jets are required to have $\pt > 20$~\GeV\ and $|\eta| < 4.8$ whilst signal jets are selected after resolving overlaps with electrons and muons, as described below, and must pass tighter requirements of $\pt > 30$~\GeV\ and $|\eta| < 2.8$.

Signal jets are identified as \bjets if they are within $|\eta| < 2.5$ and are tagged by a multivariate algorithm which uses a selection of inputs including information about the impact parameters of inner-detector tracks, the presence of displaced secondary vertices and the reconstructed flight paths of $b$- and $c$-hadrons inside the jet~\cite{Aad:2019aic}.
The \btagging algorithm used has an efficiency of 77\%, determined in a sample of simulated \ttbar events. It was chosen as part of the optimisation procedure and the corresponding misidentification rate is 20\% for $c$-jets and 0.9\% for light-flavour jets. To compensate for differences between data and MC simulation in the \btagging efficiencies and mis-tag rates, correction factors are derived from data and applied to the samples of simulated events; details are found in Ref.~\cite{Aad:2019aic}.

Electron candidates are reconstructed from energy clusters in the electromagnetic calorimeter matched to a track
in the  inner detector and are required to satisfy a set of `loose' quality
criteria~\cite{Aad:2019tso}. They are also required to lie within
the fiducial volume $|\eta| < 2.47$ and have $\pt > 4.5$~\GeV. Muon candidates are reconstructed by matching tracks in the inner detector
with tracks in the muon spectrometer. Muon candidates which have a transverse (longitudinal)
impact parameter relative to the primary vertex larger than 0.2~mm (1~mm) are rejected to suppress muons from cosmic rays.
Muon candidates  are also required to satisfy `medium' quality criteria~\cite{PERF-2015-10} and have $|\eta | < 2.5$ and $\pt > 4$~\GeV.
Electron (muon) candidates are matched to the primary vertex by requiring the transverse impact parameter ($d_0$) to
satisfy $|d_0|/\sigma(d_0) < $~5~(3), and the longitudinal impact parameter ($z_0$) to satisfy $|z_0 \sin \theta| < $~0.5~mm.
Lepton candidates remaining after resolving overlaps with baseline jets are called `baseline' leptons. In the control regions where tighter lepton identification is required, `signal' leptons are chosen from the baseline set
with \pt$ > $~20~\GeV\ and are required to be isolated from other activity in the detector using a criterion designed to accept at least 95\% of leptons from $Z$ boson decays; details are found in Ref.~\cite{TOPQ-2015-09}. In the dilepton control region where the single-lepton triggers are used, the leading lepton is required to have \pt$ > $~27~\GeV; which ensures full efficiency of the single-lepton triggers. Signal electrons are further required to satisfy `tight' quality criteria~\cite{Aad:2019tso}. The MC events are corrected to account for differences in the lepton trigger,
reconstruction and identification efficiencies between data and MC simulation.

Possible reconstruction ambiguities between baseline electrons, muons and jets are resolved by firstly removing electron candidates which share an inner detector track with a muon candidate. Jet candidates are then removed if they are within $\Delta R=\sqrt{(\Delta y)^2 + (\Delta \phi)^2} = 0.2$ of an electron candidate; next, electron candidates are discarded if they are within $\Delta R = 0.4$ of a jet. Muons are discarded if they lie within $\Delta R = 0.4$ of any remaining jet, except for the case where the number of tracks associated with the jet is less than three, where the muon is kept and the jet is discarded.
 
Identified $\tau$ leptons decaying hadronically are not considered but the following $\tau$-veto procedure is applied to reject events which contain $\tau$-like objects. Candidates ($\tau_{\mathrm{cand}}$) are identified as jets which have $|\eta| < 2.5$ and less than five inner detector tracks of $\pt > 500$~\MeV. If an event contains a tau candidate with a small azimuthal distance to the \ptmiss ($\Delta \phi (\etmiss, \tau_{\mathrm{cand}}) < \pi/3$), then the event is vetoed.
 
The missing transverse momentum $\ptmiss$ is defined as the negative vector sum of the \pt of all selected and calibrated physics objects (electrons, muons, photons~\cite{Aad:2019tso} and jets) in the event, with an extra term added to account for soft energy in the event which is not associated with any of the selected objects~\cite{ATLAS-CONF-2018-023}. This soft term is calculated from inner-detector tracks with \pt above 500~\MeV\ matched to the PV, thus ensuring it is robust against pile-up contamination~\cite{PERF-2014-04,PERF-2011-03}.

\section{Analysis strategy}
\label{sec:event_selection}
Three sets of non-orthogonal signal regions (SRs) are defined to target different mass hierarchies of the SUSY particles involved.  These definitions exploit various discriminating observables and algorithms developed to explicitly reconstruct Higgs boson candidates in
the decay chain. Events with charged leptons are vetoed in all SRs. Events with one or two charged leptons are used to define control regions (CRs) to aid in the estimation of the main SM backgrounds. Additionally, events with zero charged leptons are utilised to define validation regions (VRs) to ensure the background estimation method, described in Section~\ref{sec:background_estimation}, is robust.
The optimisation procedure for the event selection aims to maximise the yield of bottom-squark pair production events while reducing SM background contributions. It is performed for the two simplified model scenarios introduced in Section~\ref{sec:intro}. Since the $h \rightarrow bb$ decay mode is considered, the final state contains a large jet multiplicity, with many of these jets originating from $b$-quarks, and large \etmiss from the neutralinos.

The event selection criteria are defined on the basis of kinematic requirements for the objects described in the previous
section and the event variables described below. For these definitions, signal jets are used and are ordered according to decreasing \pt.
\begin{itemize}
\item $\njets$: the number of signal jets. 
\item $\nbjets$: the number of \bjets.
\item \dphiminfour: the minimum azimuthal distance between the four highest-\pt jets and the \ptmiss. This is a powerful
discriminating variable against multijet background events containing a large amount of \met due to mismeasured jets.
Typically, multijet background events exhibit low values of this variable and studies using data-driven multijet
estimates indicate that a selection of $\dphiminfour > 0.4$ is sufficient to reduce the multijet
background to a negligible level.
\item $\Delta\phi(j_1,\ptmiss)$ : the azimuthal distance between the highest-\pt jet and the \ptmiss.
This variable is used to select events where the \ptmiss is expected to be recoiling against the leading jet.
\item \meff: the effective mass~\cite{Hinchliffe:1996iu} of an event is defined as the scalar sum of the \pt of all signal jets and the \met, i.e.:
\begin{equation*}
\meff = \sum_{i\leq \njets} (p^{\mathrm{jet}}_{\mathrm{T}})_{i} + \met.
\label{eqn:Meff}
\end{equation*}
\item \metsigObj: referred to as the ``object-based \met-significance''~\cite{ATLAS-CONF-2018-038} is defined as follows:
 
\begin{equation*}
\metsigObj = \sqrt{\frac{|\ptmiss|^2}{\sigma_{\textup{L}}^2(1- \rho_{\textup{LT}}^2)}} .
\label{eqn:ObjMETSig}
\end{equation*}
 
The total momentum resolution of all jets and leptons, at a given \pt and $|\eta|$, is determined from parameterised Monte Carlo simulation which well reproduces the resolution measured in data. $\sigma_{\textup{L}}$ is the total momentum resolution after being rotated into the longitudinal (parallel to the \ptmiss) plane.  The quantity $\rho_{\textup{LT}}$ is a correlation factor between the longitudinal and transverse momentum resolution (again with respect to the \ptmiss) of each jet or lepton. The significance \metsigObj is used to discriminate events where the \etmiss arises from invisible particles in the final state from events where the \etmiss arises from poorly measured particles (and jets). Additionally, it is useful in discriminating between signal events with large \etmiss and \Zjets events with medium-to-low \etmiss.
 
\end{itemize}

Additional selections on the \pt~of the leading jet and of the leading $b$-jet are also applied as detailed in the
following subsections. In all signal regions, events containing baseline leptons with $\pt > 10$~\GeV\ are vetoed,
as well as events containing $\tau$-lepton candidates that align with the \ptmiss within $\Delta\phi = \pi/3$ . Only events with \etmiss$ > 250$~\GeV\ are retained to ensure full efficiency
of the trigger.
 
The event kinematics targeted by the three SRs are depicted in Figure~\ref{fig:Kinematic_Cartoons}.
The first signal region is SRA, designed to target the `bulk' region of both signal models, with moderate- to high-mass splitting between the \sbot and \neuttwo. In these scenarios all of the \bjets, from both the bottom-squark and Higgs boson decays, are at a relatively high \pt and can be resolved in the detector. The \bjets from the Higgs boson can be isolated by removing the ones most likely from the bottom-squark decays and checking the angular separation between the remaining \bjets.
 
The second region, SRB, is designed to target the phase space of the $\Delta m(\neuttwo, \neut) = 130$\,GeV scenario with a small mass splitting between the \sbot and \neuttwo, referred to as the ``compressed'' region. An initial-state radiation (ISR)-like selection is used where the small mass splitting between the bottom squark and neutralino leads to relatively soft \bjets from the bottom squark decay, which are difficult to reconstruct. In this scenario it is possible to reconstruct both Higgs bosons using angular separation methods. Finally, SRC is designed to target the ``compressed'' region of the $m(\neut) = 60$\,GeV signal scenario, where the mass splitting between the \sbot and \neuttwo is small. The \bjets from the bottom squark decay are very soft and as such a lower \bjet multiplicity is used in this region, when compared to the A- and B-type selections. Additionally, the visible system (\bjets from the bottom squark decay and Higgs boson decay) is produced back-to-back with the reconstructed \ptmiss.
 
\begin{figure}[t]
\begin{center}
\subfloat[SRA]{\includegraphics[width=0.3\textwidth]{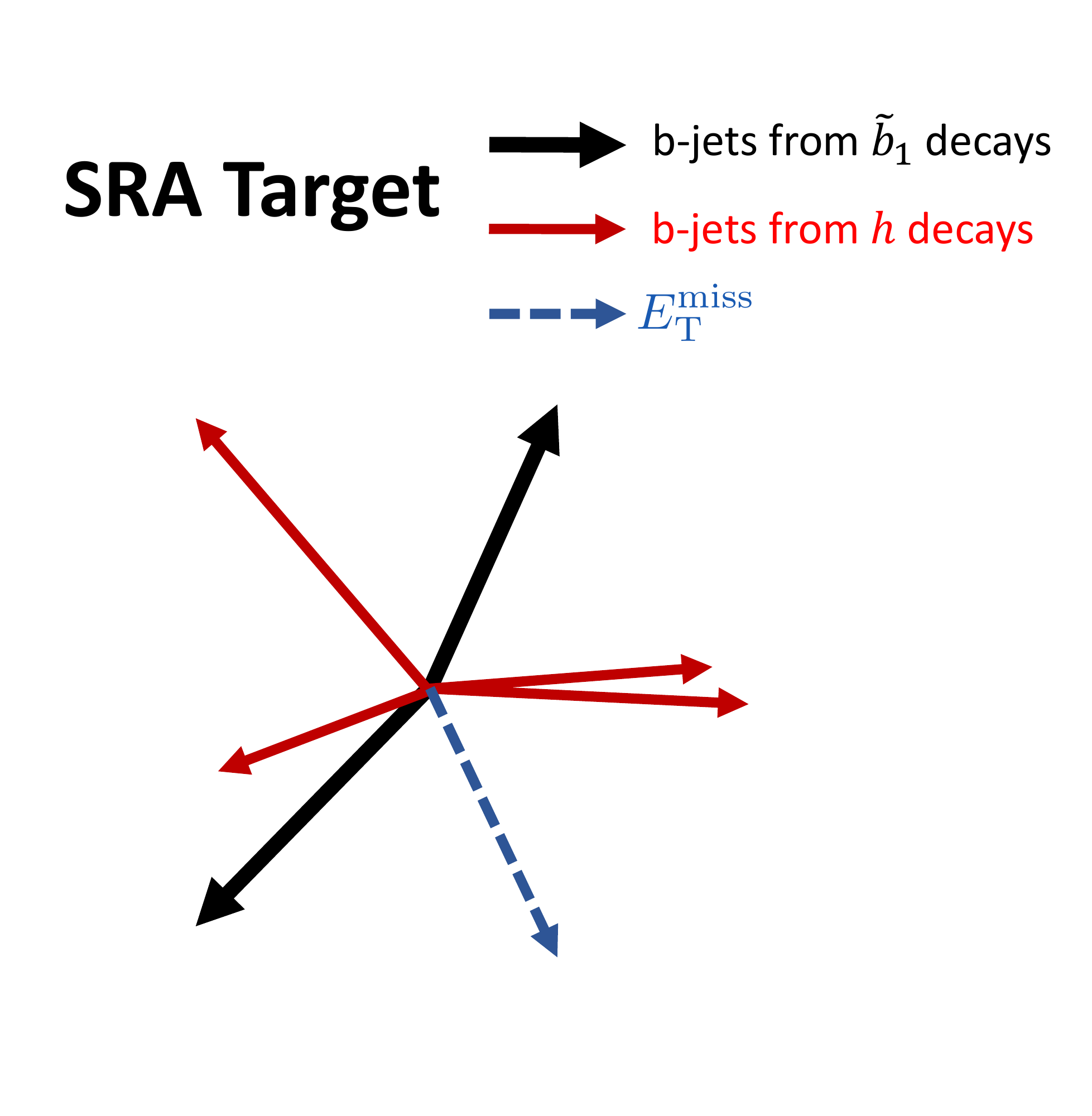}}\hfill
\subfloat[SRB]{\includegraphics[width=0.3\textwidth]{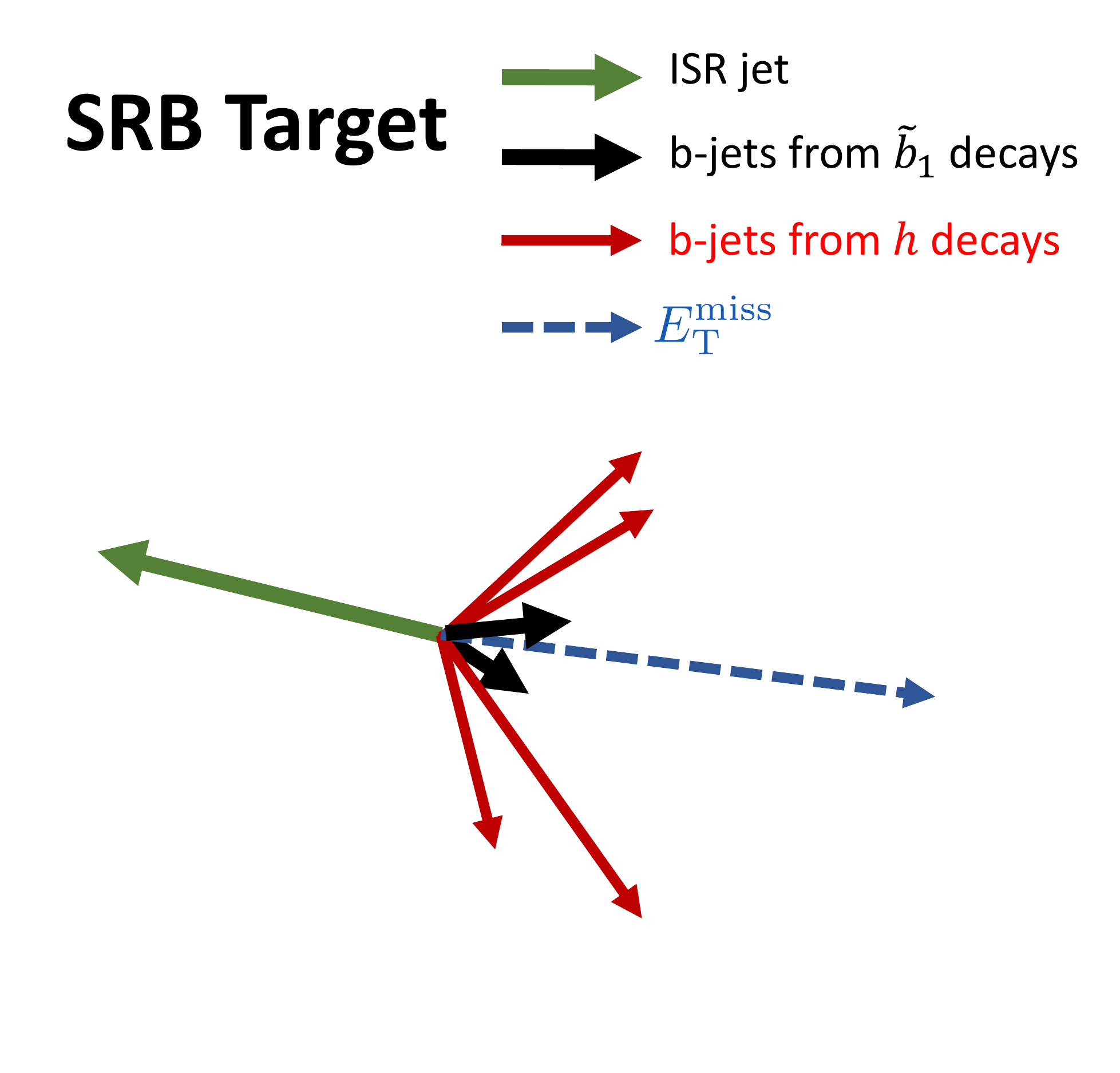}}\hfill
\subfloat[SRC]{\includegraphics[width=0.3\textwidth]{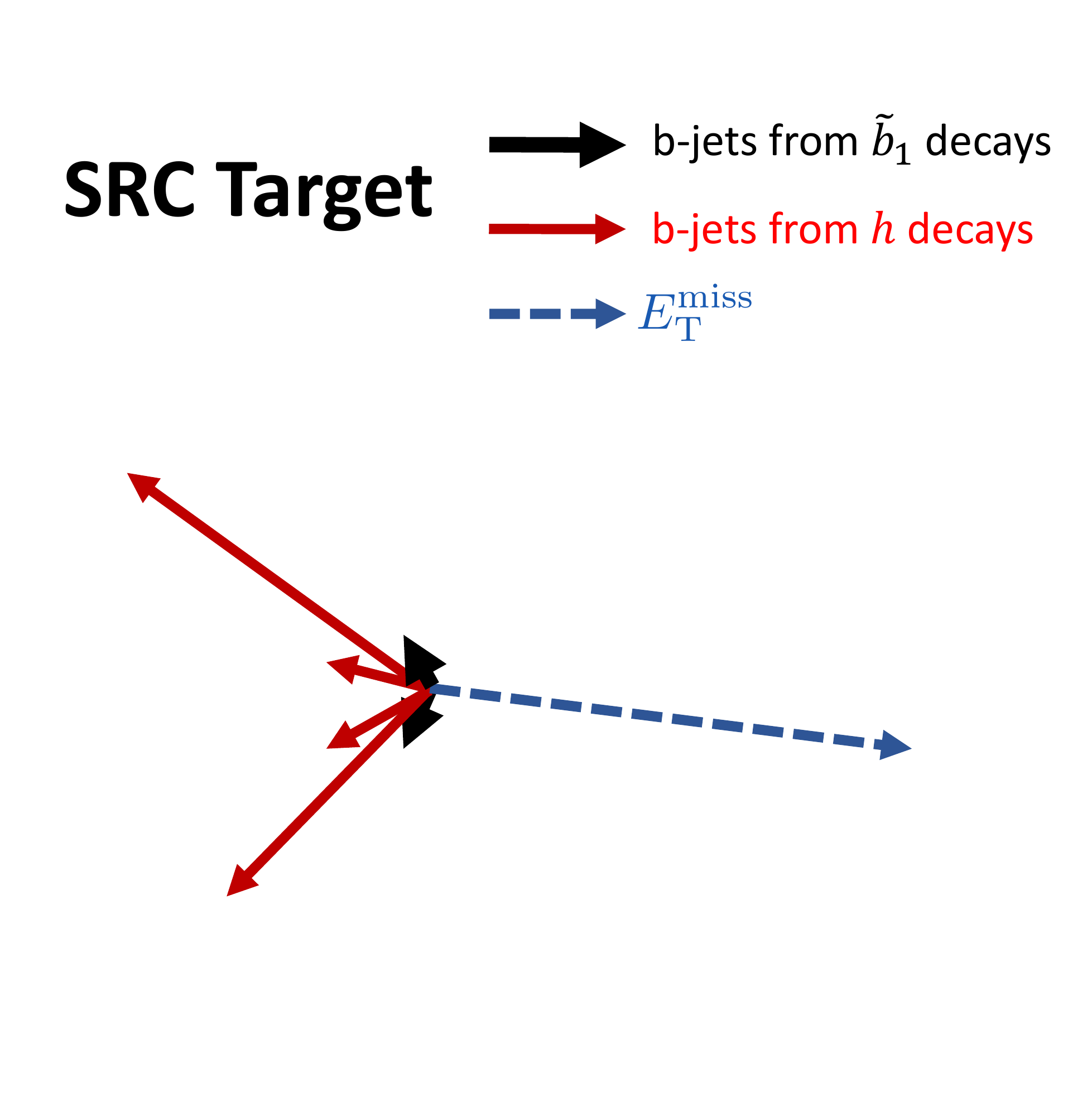}}
\end{center}
\caption{The different event kinematics, in the transverse plane, targeted by the three SRs: (a) kinematics in the bulk region, with high-\pt \bjets arising from the bottom-squark decay; (b) kinematics in the compressed region of the $\Delta m(\neuttwo, \neut) = 130$~\GeV\ scenario with soft \bjets from the bottom squark; (c) kinematics in the compressed region of the $m(\neut) = 60$~\GeV\ scenario which also contains soft \bjets from the bottom squark.}
\label{fig:Kinematic_Cartoons}
\end{figure}
 
\subsection{The SRA selections}
To exploit the kinematic properties of the signal over a large range of $\bone$, $\neuttwo$ and $\neut$ masses,
incremental thresholds are imposed on the main discriminating variable, \meff, resulting in three mutually exclusive regions, 1.0~$<$ \meff$<$~1.5~\TeV, 1.5~$<$ \meff$<$~2.0~\TeV\ and $\meff >$~2.0~\TeV. These are labelled as SRA-L, -M and -H, respectively, to maximise coverage across the \sbot mass range. The selection criteria for the three SRAs are summarised in
Table~\ref{tab:SRA_Selections}.
 
At least four \btagged jets are required. To discriminate against multijet background, events where the \ptmiss is aligned with a jet in the transverse plane are rejected by requiring $\dphiminfour > 0.4$. As a large \etmiss is expected from the neutralinos which escape the detector, a selection of $\etmiss > 350$~\GeV\ is used. Additionally, the leading \bjet ($b_{1}$) is expected to have a large \pt, hence a selection of $p_{\mathrm{T}}(b_1) > 200$~\GeV\ is employed.
At least one of the two Higgs boson candidates in the event is identified using a reconstruction algorithm referred to as \emph{max-min}, which is a two-step procedure to remove the high-\pt \bjets from the bottom squark decay and then use the remaining \bjets to reconstruct a Higgs boson in the decay chain. The procedure is implemented as follows: first, pairs of \bjets are formed by iterating through all of the \bjets in the event, and the pair with the largest separation in $\Delta R$ is designated as arising from the bottom-squark decay and removed from the subsequent step; second, the pair with the smallest $\Delta R$ is identified as a possible Higgs boson candidate and its invariant mass calculated.
The following $\Delta R$ and mass quantities are defined:
\begin{itemize}
\item \maxDRbb: the distance in $\eta$--$\phi$ between the two $b$-jets with the maximal angular separation
which are most likely to originate from the initial decay of the $\bone$;
\item \maxminDRbb: the distance in $\eta$--$\phi$ between the two $b$-jets with the minimum angular separation which are most likely to originate from the same Higgs boson decay, selected out of the remaining \bjets;
\item \maxminDRmbb: the invariant mass of the $b$-jet pair identified as a Higgs candidate by the \emph{max-min} algorithm. A lower bound on \maxminDRmbb is used; in the majority of events the distribution peaks around the Higgs boson mass, but in scenarios where the incorrect combination of \bjets is chosen the signal can extend to higher masses.
\end{itemize}
 
When applied to signal, the \emph{max-min} algorithm correctly selects a $h \rightarrow bb$ pairing in 20\%--40\% of cases for a single Higgs boson decay, depending upon the model. For a signal model corresponding to $m(\sbot, \neuttwo, \neut) = (1100, 330, 200)$~\GeV, about 3\% of the simulated signal events are retained by the SRA selections. 
 
\begin{table}[!htb]
\caption{Definitions for the SRA, alongside the three varying \meff intervals used. The letter appended to the SRA label corresponds to the low (-L), medium (-M) or high (-H) \meff selection. This selection is sensitive to the bulk regions of both signal scenarios. The jets and \bjets are ordered by \pt.}
\label{tab:SRA_Selections}
\centering
\begin{tabular}{lc|ccc}
\toprule
Variable &  SRA & SRA-L & SRA-M & SRA-H \\
\midrule
\nlep(baseline) & = 0 & \multicolumn{3}{c}{= 0} \\
\njets & $\geq 6$& \multicolumn{3}{c}{$\geq 6$} \\
\nbjets & $\geq 4$& \multicolumn{3}{c}{$\geq 4$} \\
\met [\GeV] & $ > 350$ & \multicolumn{3}{c}{$ > 350$} \\
\dphiminfour [rad] &$> 0.4$ & \multicolumn{3}{c}{$> 0.4$ }\\
$\tau$ veto & Yes &\multicolumn{3}{c}{Yes} \\
$p_{\mathrm{T}}(b_1)$ [\GeV] & $> 200$ & \multicolumn{3}{c}{$> 200$}\\
\maxDRbb & $> 2.5$ &\multicolumn{3}{c}{$> 2.5$ } \\
\maxminDRbb & $< 2.5$ &\multicolumn{3}{c}{$< 2.5$}\\
\maxminDRmbb [\GeV] & $> $ 80 &\multicolumn{3}{c}{$> $ 80}\\
\meff [\TeV] & $> 1.0$ &  $\in[1.0 , 1.5] $ &  $\in[1.5 , 2.0]$ &  $>2.0$ \\
\bottomrule
\end{tabular}
\end{table}
 
\subsection{The SRB selections}
The SRB region targets small mass-splitting between the \bone and \neuttwo (of order 5--20~\GeV), in the case of the $\Delta m (\neuttwo, \neut) = 130$~\GeV\ scenarios. The presence of an ISR jet boosting the bottom squarks, and consequently
their decay products, is exploited.
To efficiently suppress SM background contributions, events are selected where the highest-\pt jet is not \btagged and has \pt $>$~350~\GeV; this jet is presumed to arise from ISR in the scenario under consideration. Additional selections of \etmiss$ > 350$~\GeV\ and $\Delta\phi(j_1,\met) > 2.8$ are applied. An \meff selection of $>$~1~\TeV\ is also applied.
The soft \pt spectrum predicted for \bjets from \bone decays can cause the \bjets to be difficult to reconstruct, hence a different algorithm, aiming to reconstruct both Higgs boson candidates, is employed.

Differently from the scenarios targeted by SRA, pairs of \bjets with the largest $\Delta R$ are found to be more
likely to arise from the decay of the same Higgs boson candidate.
Two pairs at a time are identified following an iterative procedure, such that at first the pair of \bjets leading to the highest $\Delta R$,  $\Delta R_{bb1}$, is defined, followed by the second highest $\Delta R$, $\Delta R_{bb2}$,
built considering only the remaining \bjets.
The average mass of the two candidates $m(h_{\mathrm{cand1}},h_{\mathrm{cand2}})_{\mathrm{avg}}$
is calculated and a requirement is placed on the average mass, corresponding to a window around
the Higgs boson mass: [75, 175]~\GeV. For a signal model corresponding to $m(\sbot, \neuttwo, \neut) = (700, 680, 550)$~\GeV, about 0.1\% of the simulated signal events are retained by the SRB selections. The efficiency of correctly selecting the \bjets using this algorithm is in the range 15\%--30\%.
The SRB requirements are listed in Table~\ref{table:SRB_Selections}.
 
\begin{table}[htb!]
\caption{Definitions for SRB, targeting the compressed region of the $\Delta m(\neuttwo, \neut) = 130$~\GeV\ scenario. The jets and \bjets are ordered by \pt.}
\label{table:SRB_Selections}
\centering
\begin{tabular}{lc}
\toprule
Variable & SRB \\
\midrule
\nlep(baseline) & = 0 \\
\njets & $\geq 5$ \\
\nbjets & $ \geq 4 $\\
\met [\GeV] & $ > 350$ \\
\dphiminfour [rad] & $>$ 0.4\\
$\tau$ veto & Yes \\
$m(h_{\mathrm{cand1}},h_{\mathrm{cand2}})_{\mathrm{avg}}$ [\GeV] & $\in [75,175]$ \\
Leading jet not \btagged & Yes \\
$p_{\mathrm{T}}(j_1)$ [\GeV] & $>350$ \\
$|\Delta\phi(j_1, \met)|$ [rad] & $>$ 2.8\\
\meff [\TeV] & $> 1$ \\
\bottomrule
\end{tabular}
\end{table}

\subsection{The SRC selections}
When considering the scenario with a constant \neut mass of 60~\GeV, the $\Delta R$-based Higgs boson reconstruction algorithms are ineffective in the compressed region of phase space with a small mass splitting between the \sbot and \neuttwo. 
In the inclusive SRC, the main discriminating quantity is \metsigObj; a selection of $\metsigObj > 22$ is employed.
Events are also required to have at least three \bjets. Four non-overlapping regions (SRC22, SRC24, SRC26 and SRC28) are defined as subsets of the inclusive SRC region, with incremental thresholds placed on \metsigObj as detailed in Table \ref{tab:SRC_Selections}, to ensure full coverage of the target models as a function of bottom-squark and neutralino mass. For a signal model corresponding to $m(\sbot, \neuttwo, \neut) = (1200, 1150, 60)$~\GeV, about 11\% of the simulated signal events are retained by the SRC selections. The \metsigObj variable is effective in rejecting the SM background arising from associated production of a $Z$ boson decaying into neutrinos and \bjets.
 
\begin{table}[h]
\caption{Definitions for SRC, alongside the four varying \metsigObj~intervals used. The letter appended to the SRC label corresponds to the lower bound on the \metsigObj~interval. SRC targets small mass splittings between the \sbot and \neuttwo, in the $m(\neut) = 60$~\GeV\ signal scenario. The jets and \bjets are ordered by \pt.}
\label{tab:SRC_Selections}
\centering
\begin{tabular}{lc|cccc}
\toprule
Variable & SRC & SRC22 & SRC24 & SRC26  & SRC28  \\
\midrule
\nlep(baseline) & = 0 & \multicolumn{4}{c}{ = 0 } \\
\njets   & $\geq 4$ & \multicolumn{4}{c}{$\geq 4$} \\
\nbjets    & $\geq 3$  &  \multicolumn{4}{c}{$\geq 3$} \\
\met       [\GeV] & $ > 250$  & \multicolumn{4}{c}{$ > 250$} \\
\dphiminfour [rad] & $> 0.4$  & \multicolumn{4}{c}{$> 0.4$} \\
\metsigObj &   $> 22$ & $\in[22,24]$  & $\in[24,26]$  & $\in[26,28]$  & $> 28$\\
\bottomrule
\end{tabular}
\end{table}

 
\section{Background estimation}
\label{sec:background_estimation}
There are two main SM backgrounds which are expected to contribute to the yields for the SRs introduced in the previous section. For SRAs and SRB, the main background is top-quark production which, according to MC estimates,  contributes between 70\% and 85\% of the total background, depending upon the region considered, and is dominated by top-quark pairs produced in association with two \bquarks arising from gluon splitting. In the SRCs, the main backgrounds arise from \Zjets (up to 50\% of the total) and from top-quark-related processes (up to 20\% of the total).
 
The main SM backgrounds in each SR are determined separately with a profile likelihood fit to the event yields in the associated CRs~\cite{Baak:2014wma}. This is commonly referred to as a background-only fit which constrains and adjusts the normalisation of the background processes. The background-only fit uses the observed event yield and the expected number of MC events in the associated CRs, which are described by Poisson statistics, as a constraint to adjust the normalisation of the background processes assuming that no signal is present.
 
The normalisation factor is referred to as the $\mu$ factor. The CRs are designed to be enriched in specific background contributions relevant to the analysis, whilst minimising the potential signal contamination, and they are orthogonal to the SRs.
 
When performing the fit for SRA, a multi-bin approach is used, with a single CR divided into three bins of \meff. Such an approach allows the calculation and use of a single normalisation parameter (applied to the main \ttbar background across all bins of \meff), and additionally enables the fit to take into account the modelling of the \meff variable.
 
The systematic uncertainties, described in \Sect{\ref{sec:systematic_uncertainties}}, are included in the fit as nuisance parameters. They are constrained by Gaussian distributions with widths corresponding to the sizes of the uncertainties and are treated as correlated, when appropriate, between the various regions. The product of the various probability density functions and the Gaussian distributions forms the likelihood function, which the fit maximises by adjusting the background normalisation
and the nuisance parameters. This approach reduces the influence of systematic uncertainties on the backgrounds with dedicated CRs, as these are absorbed by the normalisation parameter.
 
Finally, the reliability of the MC extrapolation of the SM background estimates outside of the CRs is evaluated in dedicated VRs, orthogonal to CRs and SRs.
 
\begin{figure}[!htb]
\centering
\subfloat[]{\includegraphics[width=0.5\columnwidth]{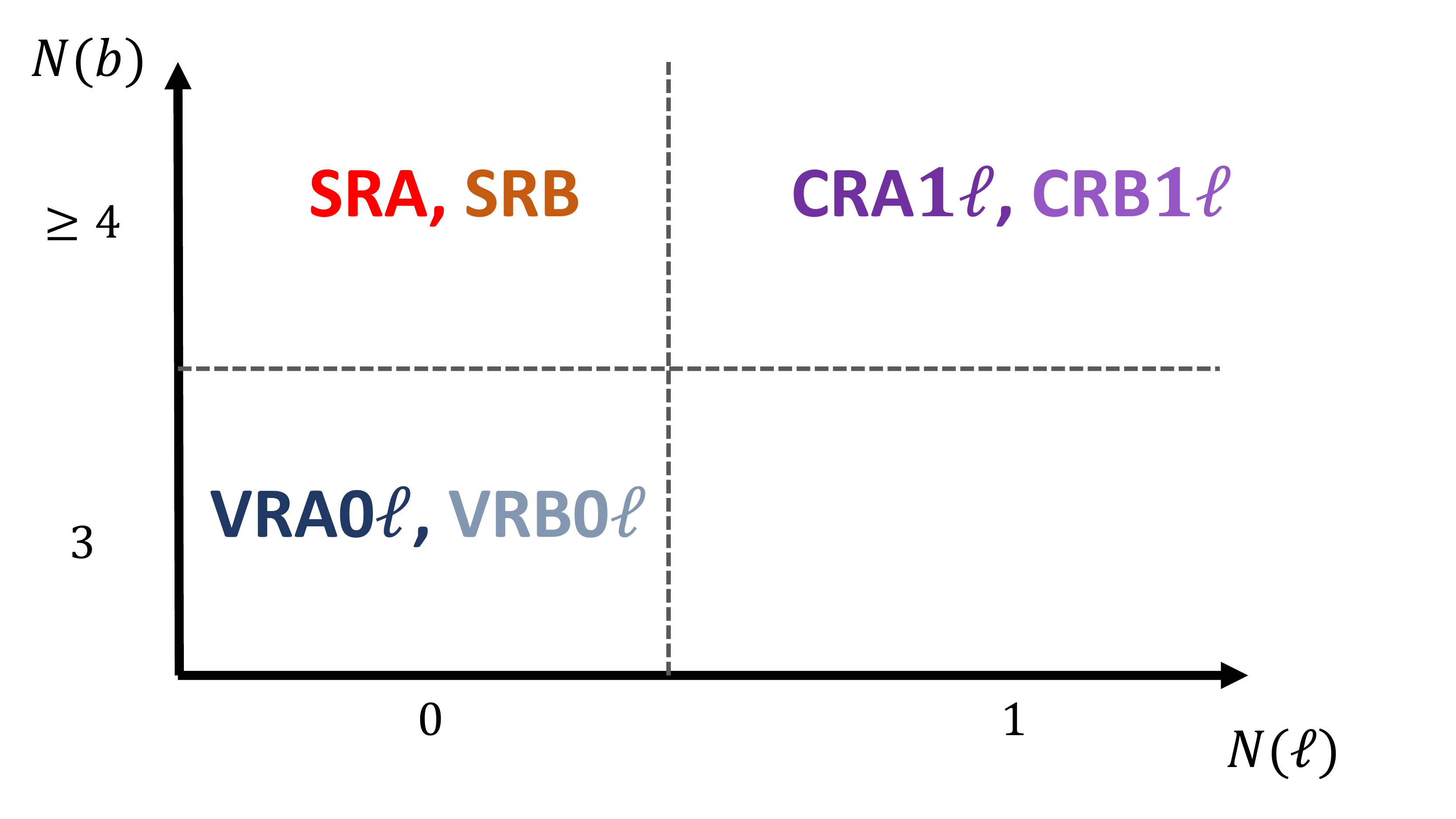}\label{fig:ABFitCartoon}}
\subfloat[]{\includegraphics[width=0.5\columnwidth]{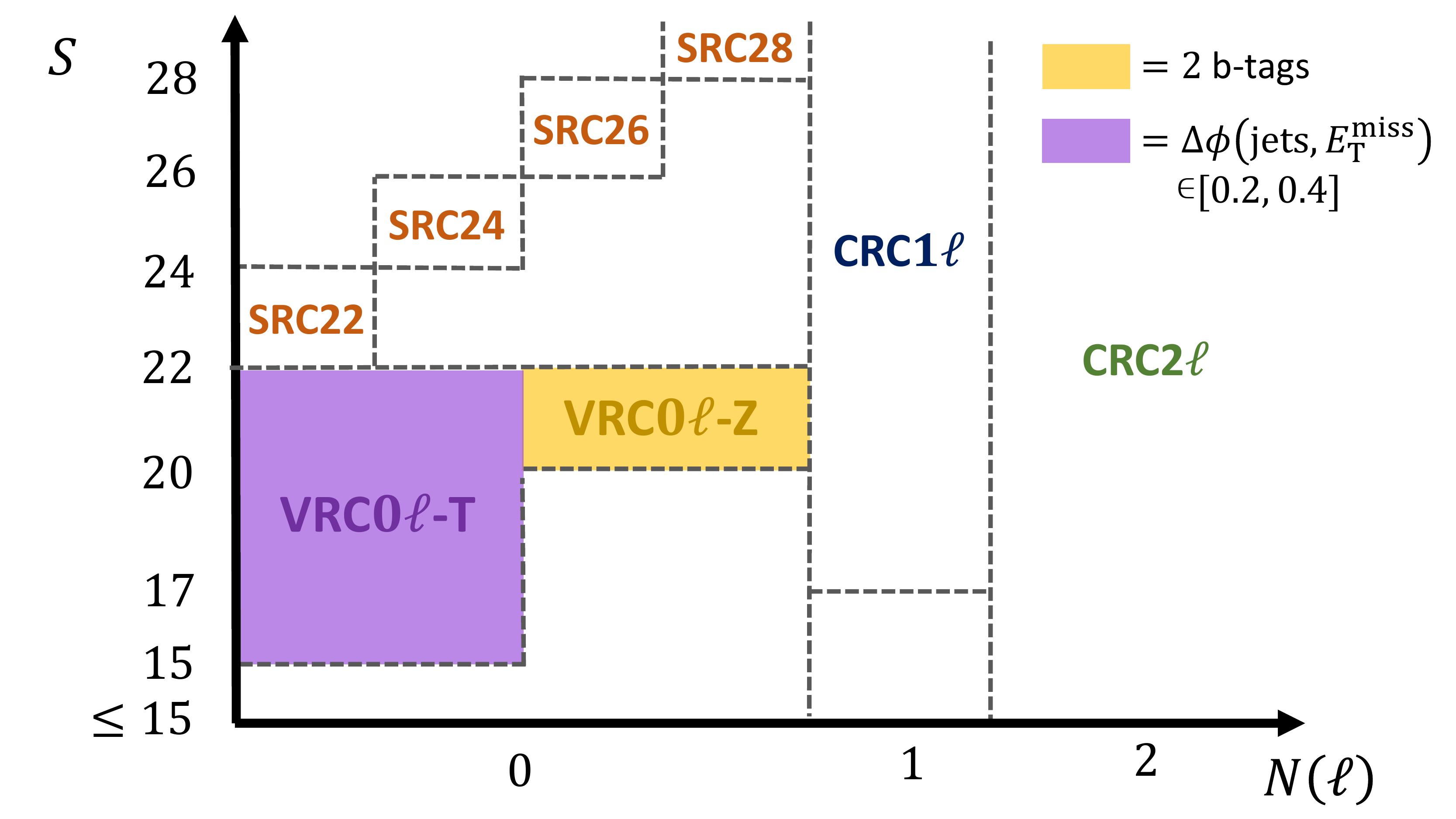}\label{fig:CFitCartoon}} \\
\caption{Schematic diagrams of the fit strategies for (a) the A-, B- and (b) C-type regions. Generally the CRs require a different lepton multiplicity than the SRs. The validation regions are defined with a lower \bjet multiplicity requirement, except in the case of the VRC0$\ell$-T region, which instead inverts the SR \dphiminfour selection. }
\label{fig:Cartoons}
\end{figure}
 
The fit strategies for the A- and B-type regions are very similar and are represented schematically in Figure~\ref{fig:ABFitCartoon}. They rely on CRs with a single-lepton requirement, as the \ttbar background in the SR is dominated by semileptonic \ttbar decays where the lepton is not identified. The main background in both regions is \ttbar pair production in association with heavy-flavour jets. The fit strategy for the C-type regions is presented in Figure~\ref{fig:CFitCartoon}. The strategy is different because the main background in these regions is $Z$+jets, closely followed by the top-quark backgrounds. In order to define CRs enhanced in \ttbar and $Z$+jets, additional variables are used:
 
\begin{itemize}
\item $\mt$ : the event transverse mass \mt is defined as $\mt = \sqrt{2 \pt(\ell) \etmiss(1 - \cos(\Delta \phi))} $, where $\Delta \phi$ is the difference in azimuthal angle between the lepton and the \ptmiss. This is used in the one-lepton CRs to reject multi-jet events which can be misidentified as containing a prompt lepton.

\item $m_{\ell\ell}$: the invariant mass of the two leptons in the event. Since the two-lepton CR is used to constrain the $Z$+jets background, the $m_{\ell\ell}$ variable is required to be within the $Z$-mass window: [86, 106]~\GeV\ (used exclusively in the two-lepton CR).
 
\item $\METprime$: the `lepton corrected' \met. For the two-lepton CR the transverse momentum vectors of the leptons are subtracted from the \met calculation in order to mimic the neutrinos from $Z \rightarrow \nu\nu$ decays (used exclusively in the two-lepton CR).
 
\end{itemize}
 
When designing the CRs and VRs, the potential signal contamination is checked in each region to ensure that the contribution from the signal process being targeted is small in the regions. The signal contamination in the CRs and VRs is found to be negligible, at the level of $<$~1\% of the total SM expectation, depending upon the signal mass hierarchy of the models considered in this search. 
 
\subsection{A-type CR and VR definitions}
 
A single, \ttbar-dominated CR (CRA1$\ell$) is defined for the A-type regions and is split into the same three identical \meff selections as the SRAs. The CR is defined similarly to the SR selection (as documented in Table~\ref{tab:SRA_Selections}); however, exactly one signal lepton (either $e$ or $\mu$) with \pt$>$~20~\GeV\ is required in the final state. Furthermore, the selections used to isolate the Higgs boson in the SRAs, namely the \maxDRbb, \maxminDRbb\ and \maxminDRmbb\ selections, are not applied in order to increase the number of events in the CR. The leading \bjet \pt selection is lowered to $> 100$~\GeV\ to further increase the number of events in the region, and a selection on the transverse mass of $\mT > 20$~\GeV\ is applied to suppress misidentified leptons. Such selections result in pure \ttbar CRs, with \ttbar contributing more than 80\% of the total SM contribution in each of the CRs. The fraction of top-quark pairs produced in association with \bquarks is equivalent between CRs and SRs, and accounts for about 70\% of the total \ttbar background.
Figure~\ref{fig:CR_higgsMasses_CRA} presents the distribution of \maxminDRmbb\ in CRA1$\ell$, and shows that this variable
is well modelled.
 
A zero-lepton validation region (VRA0$\ell$) is also defined, and split according to the same \meff thresholds as the SRAs and CRAs. This VR is used to validate the modelling of the \ttbar background when extrapolating from the one-lepton CRs to zero-lepton regions. The selections are based upon the SR selections but the VRs are orthogonal due to the \bjet multiplicity selection, which requires exactly three \bjets. Additionally, the \maxDRbb, \maxminDRbb\ and \maxminDRmbb selections are not applied in this region. A selection of \metsigObj$ <$~22 is applied to ensure this region is orthogonal to the SRC regions.
 
\subsection{B-type CR and VR definitions}
 
For the B-type \ttbar CR (CRB1$\ell$), a similar method of using a one-lepton region enriched in \ttbar is implemented. The SR selections (as documented in Table~\ref{table:SRB_Selections}) are applied, and additionally exactly one signal lepton with \pt $>$~20~\GeV\ is required. The $m(h_{\mathrm{cand1}},h_{\mathrm{cand2}})_{\mathrm{avg}}$ selection is dropped to increase the number of events in the region, and the $|\Delta\phi(j_1, \met)|$ selection is loosened to $> 2.2$. Similarly to the A-type CR, a selection of $\mT > 20$~\GeV\ is applied to suppress misidentified leptons. These selections result in a pure CR with 80\% of the total expected SM background consisting of \ttbar. Figure~\ref{fig:CR_higgsMasses_CRB} presents the $m(h_{\mathrm{cand1}},h_{\mathrm{cand2}})_{\mathrm{avg}}$ distribution in this region; it is shown to be well modelled.
 
The associated VR (VRB0$\ell$) is defined in a similar manner to the A-type VR, with selections similar to those of the SRB region, but an exclusive \bjet multiplicity selection of exactly three \bjets. Additionally, the selections used to reconstruct the Higgs bosons in the event are dropped to enhance the number of events in the region. A selection of \metsigObj$ < $~22 is also applied to ensure this region is orthogonal to the C-type SRs.
 
\begin{figure}[t]
\begin{center}
\subfloat[]{\includegraphics[width=0.5\textwidth]{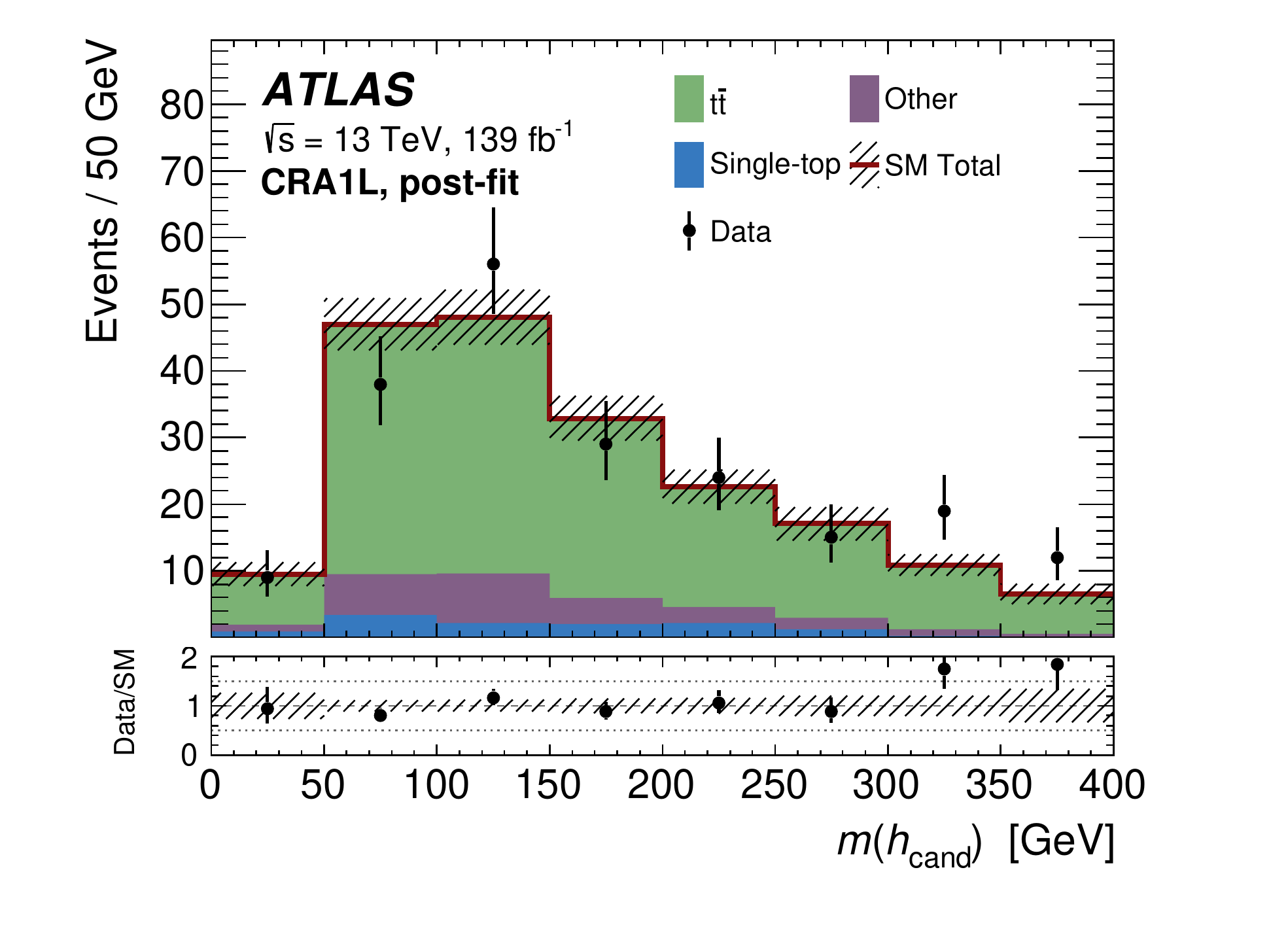}\label{fig:CR_higgsMasses_CRA}}
\subfloat[]{\includegraphics[width=0.5\textwidth]{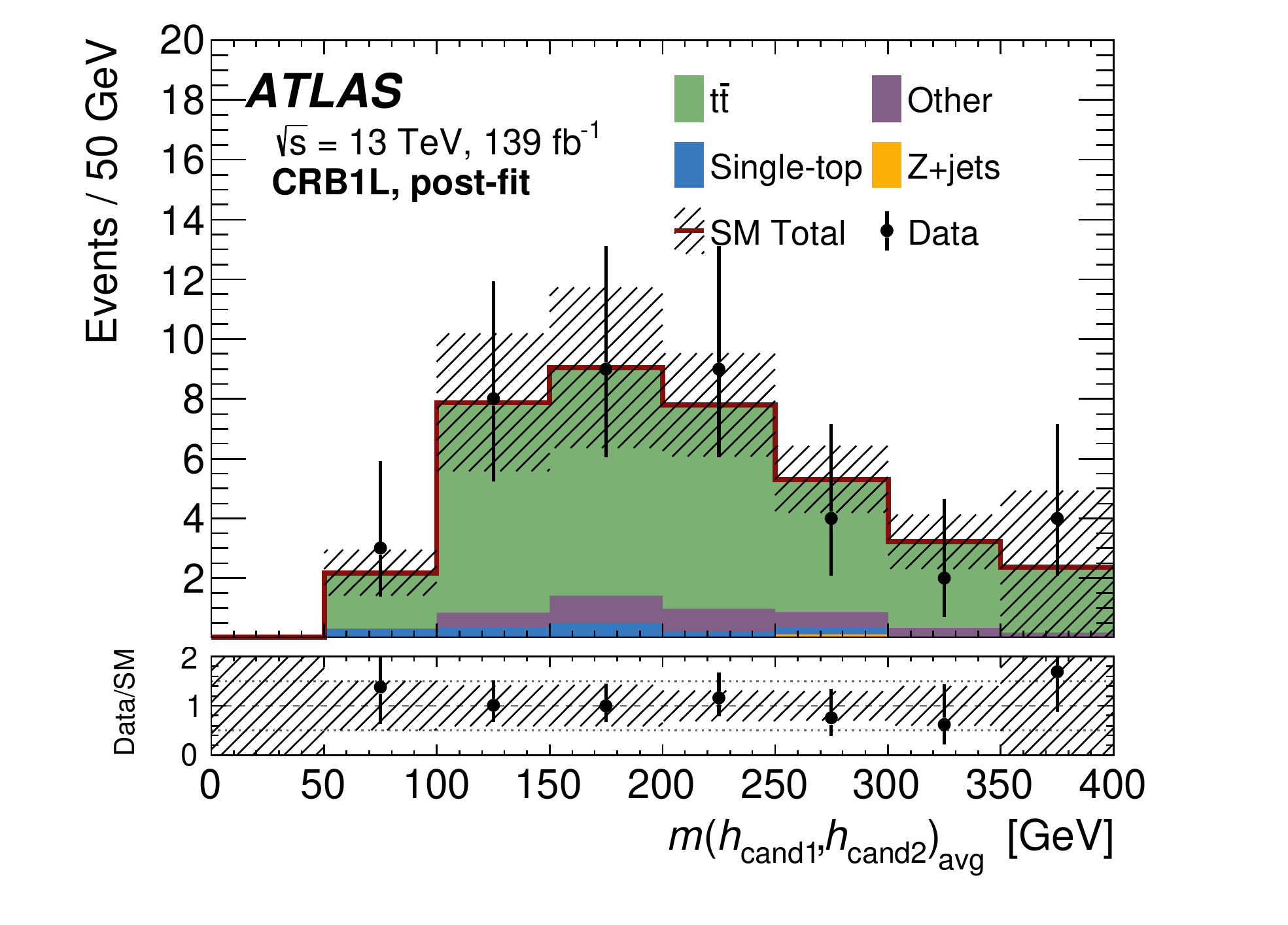}\label{fig:CR_higgsMasses_CRB}} \\
\subfloat[]{\includegraphics[width=0.5\textwidth]{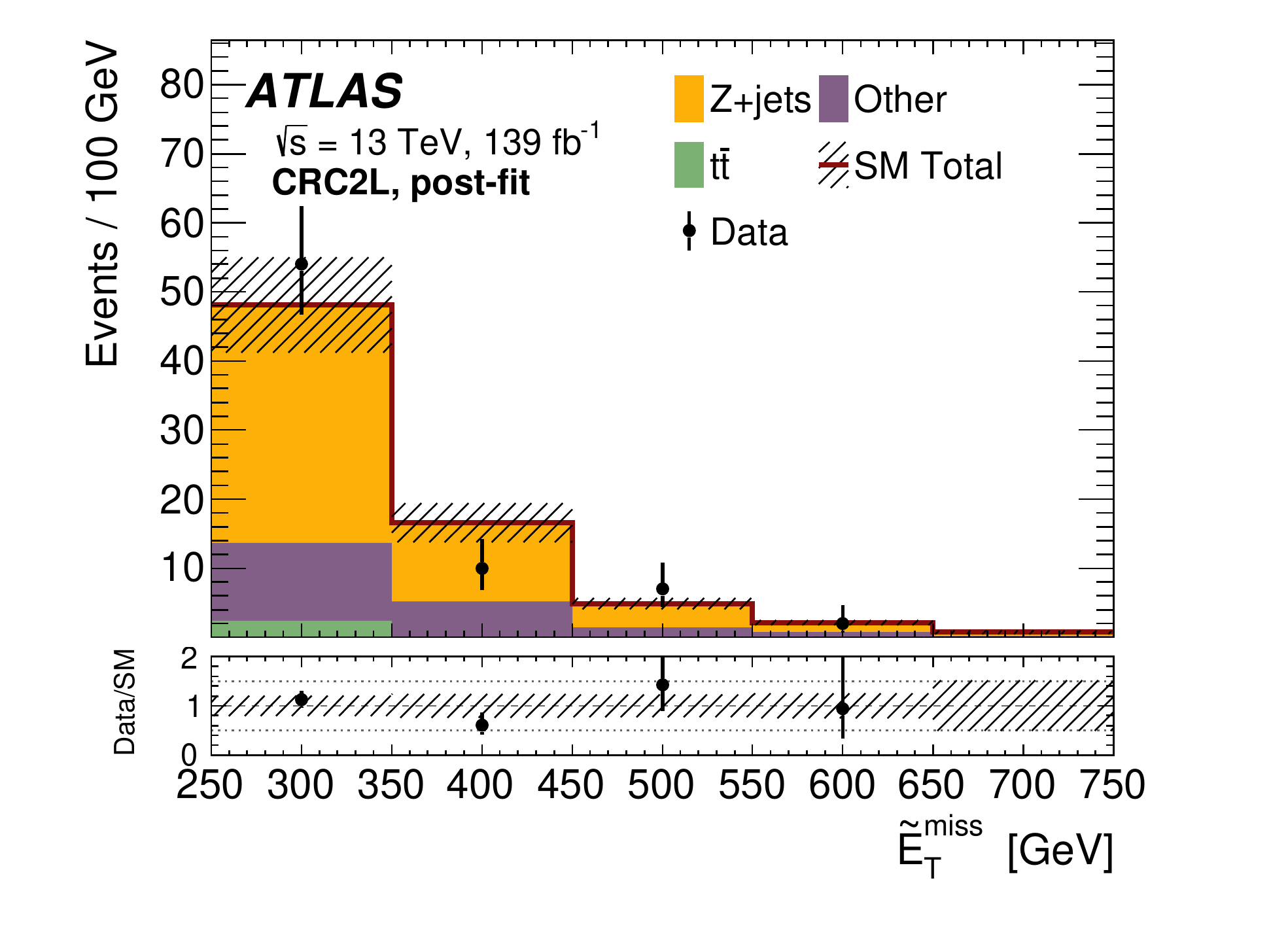}\label{fig:CR_higgsMasses_CRC2L}}
\subfloat[]{\includegraphics[width=0.5\textwidth]{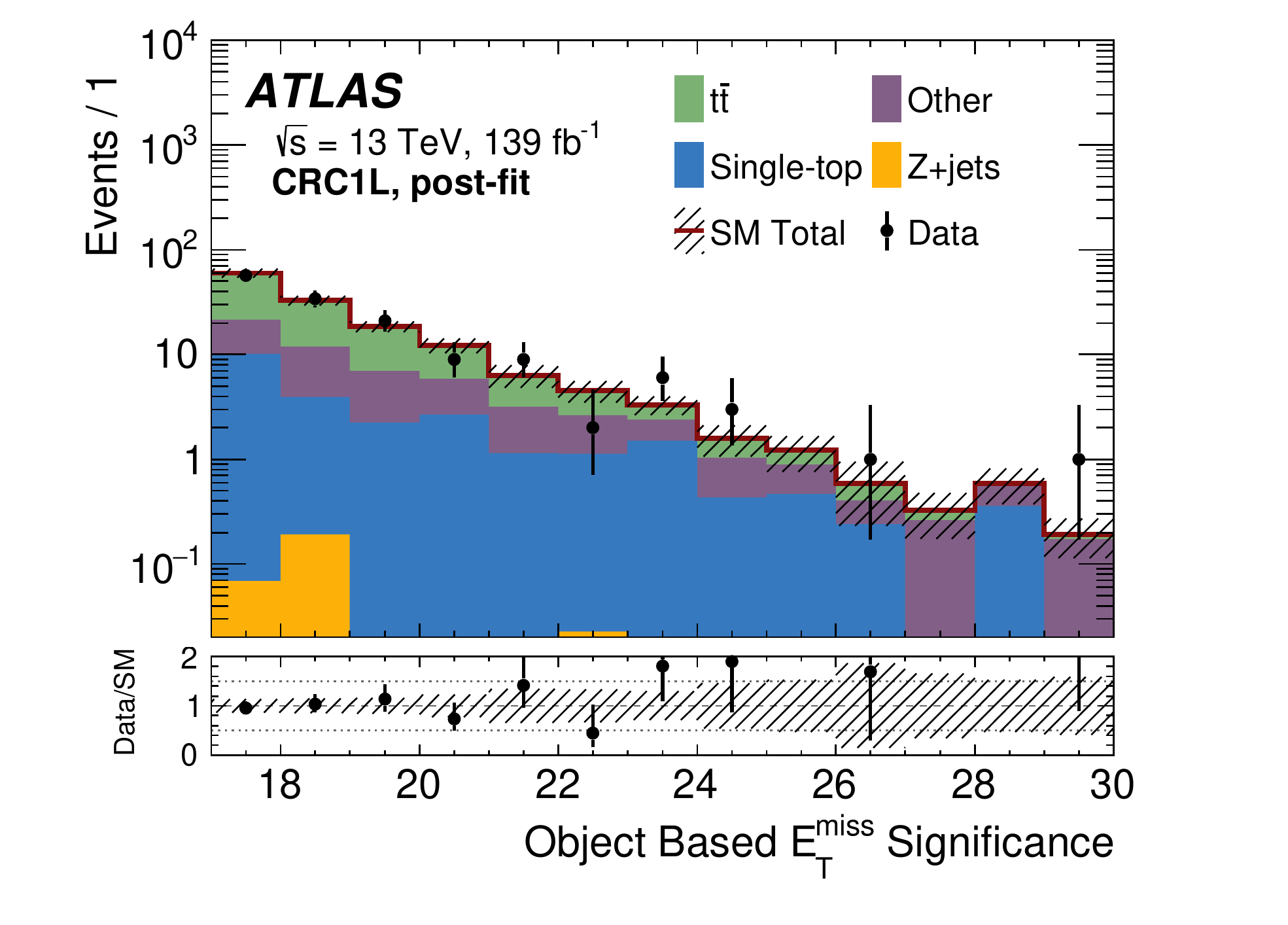}\label{fig:CR_higgsMasses_CRC}}
 
\end{center}
\caption{Distributions of (a) \maxminDRmbb in CRA1$\ell$, (b) $m(h_{\mathrm{cand1}},h_{\mathrm{cand2}})_{\mathrm{avg}}$ in CRB1$\ell$, (c) \METprime in CRC2$\ell$, and (d) \metsigObj in CRC1$\ell$  after the background-only fit; ratios of data to SM predictions are reported in the bottom panels. All uncertainties as defined in Section~\ref{sec:systematic_uncertainties} are included in the uncertainty bands of the top and bottom panels in each plot. The backgrounds which contribute only a small amount (diboson, $W$+jets and \ttbar$+W/Z/h$) are grouped and labelled as `Other'. Overflow events which do not fall into the axis range are placed into the rightmost bin.}
\label{fig:CR_higgsMasses}
\end{figure}

\subsection{C-type CR and VR definitions}
 
Two CRs are defined for the C-type SRs, one to constrain the $Z$+jets background (CRC2$\ell$) and one to constrain the backgrounds associated with top quarks, \ttbar and single top (CRC1$\ell$). A single normalisation parameter is used to constrain both the \ttbar and single-top backgrounds, while the \Zjets background is constrained with an additional normalisation parameter. These CRs are based upon the SR shown in Table~\ref{tab:SRC_Selections}, but are orthogonal due to the different lepton multiplicities required.
 
The CRC2$\ell$ requires two same-flavour (SF) opposite-sign (OS) leptons, with invariant mass in the $Z$-mass window. The leading two leptons are required to have \pt$>$~27~\GeV\ and  \pt$>$~20~\GeV\ respectively. To imitate the \met selection in the SR, a selection of $\METprime > 250$~\GeV\ is utilised. For this region the selections on \metsigObj are dropped to enhance the number of events in the region. Figure~\ref{fig:CR_higgsMasses_CRC2L} shows the \METprime distribution in this region. The CRC1$\ell$ region used to constrain the top-quark-related backgrounds requires one signal lepton with \pt$>$~20~\GeV. A selection of \metsigObj~$>$~17 is applied. Similarly to the A- and B-type CRs, a selection of \mt$>$~20~\GeV\ is applied to remove the multi-jet contribution with fake or non-prompt leptons. Figure~\ref{fig:CR_higgsMasses_CRC} presents the \metsigObj distribution in this region.
 
Two zero-lepton VRs are defined to validate the extrapolation from CR to SR based on the SR selections. A VR with zero leptons and two \bjets (VRC0$\ell$-Z) with \metsigObj $\in [20, 22]$ and \met $\in [250, 600]$~\GeV\ ensures a region orthogonal to the SR, but with a large contribution from the $Z$+jets process. A second VR is used to validate the modelling of the \ttbar and single-top backgrounds (VRC0$\ell$-T); a selection of zero leptons, \metsigObj $\in [15, 22]$ and an inverted selection on the \dphiminfour $\in [0.2, 0.4]$ is applied to ensure orthogonality.
 
\subsection{Summary of CR and VR results}
 
A full overview of the control and validation regions used in the analysis can be found in Table~\ref{tab:summaryCRVR}. The control region pre-fit yields and fitted normalisation factors $\mu_{\mathrm{bkg}}$ for the A-, B- and C-type regions are presented in Figure~\ref{fig:CR_Fit}. All $\mu$ values are consistent with unity, within 2$\sigma$ of the normalisation uncertainty, suggesting the modelling of the key SM background processes is already good before performing the fit.
Figure~\ref{fig:VR_Fit} presents the observed yields, post-fit background estimates and significance \cite{SignificanceCalc} for the A-, B- and C-type validation regions. The background-only fit estimates are
in good agreement with the data in these regions, and the post-fit expectation is within 1$\sigma$ of the central value for all regions.
 
\begin{figure}[!htb]
\centering
\subfloat[]{\includegraphics[width=0.65\textwidth]{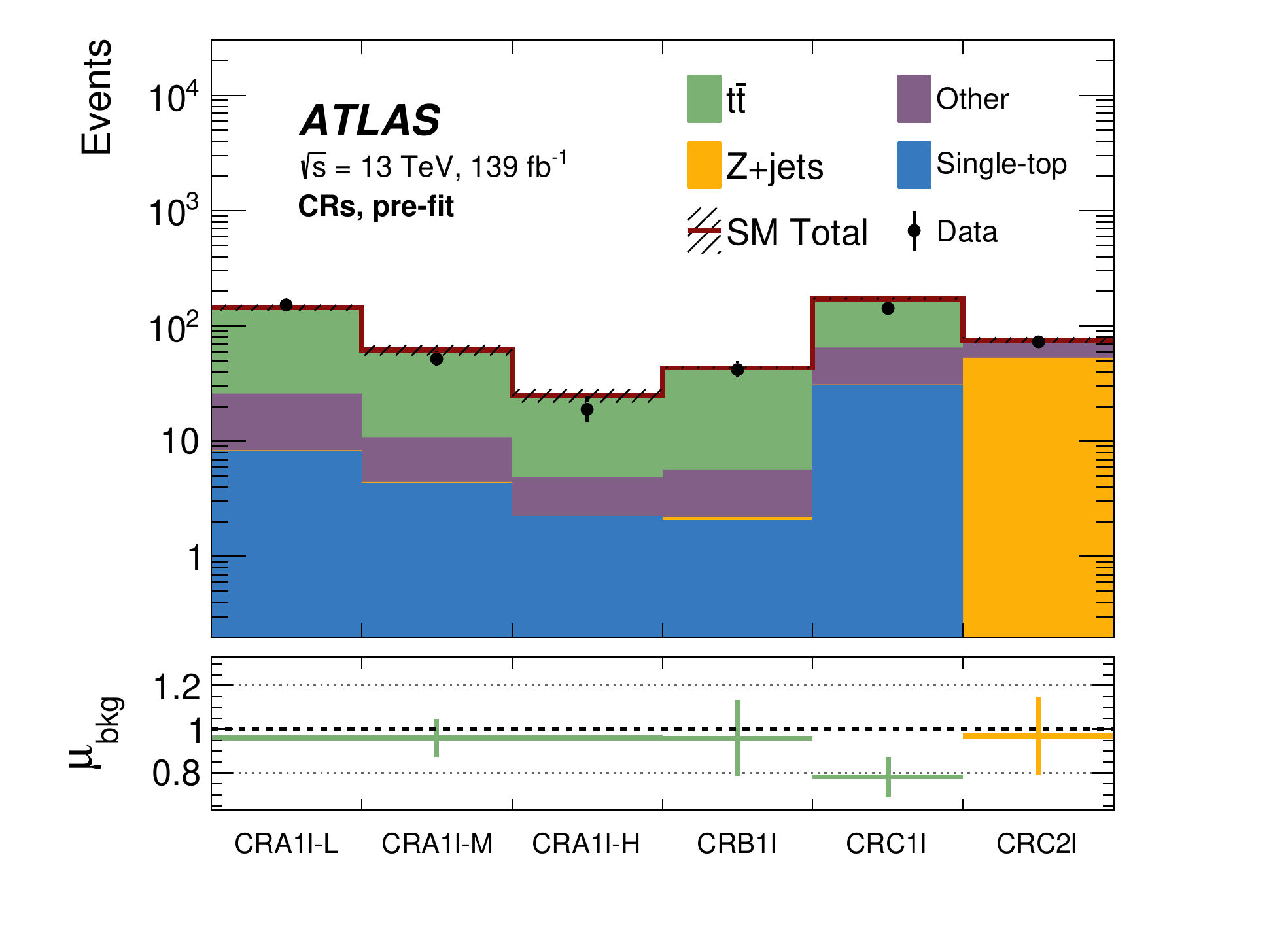}\label{fig:CR_Fit}}\\
\subfloat[]{\includegraphics[width=0.65\textwidth]{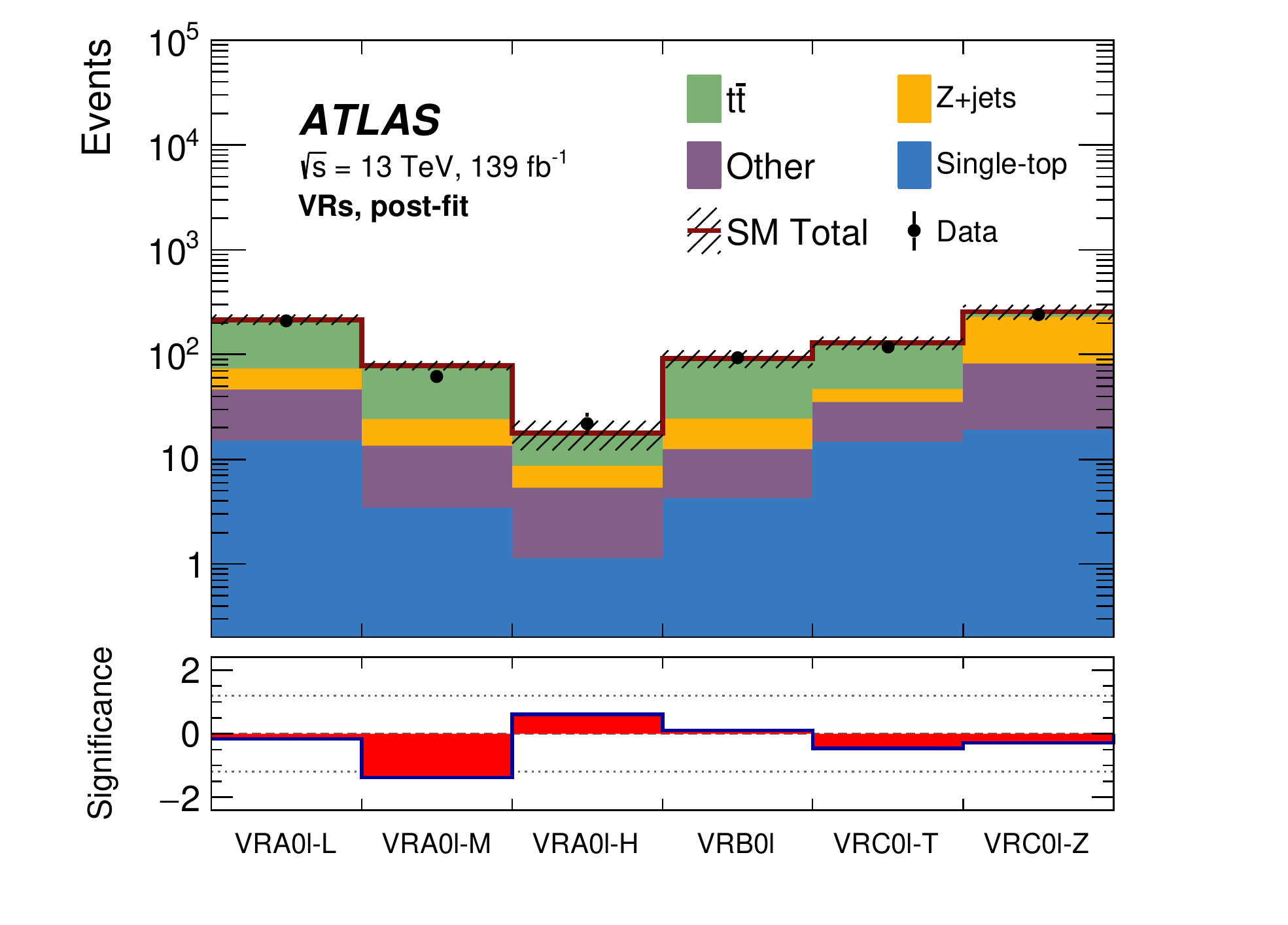}\label{fig:VR_Fit}}
\caption{(a) Control region event pre-fit event yields compared with SM MC predictions (top) and post-fit $\mu$ scale factors (bottom) for the A-, B- and C-type regions. The uncertainty in the $\mu$ factors and the total expected yield include statistical and systematic uncertainties as introduced in Section~\ref{sec:systematic_uncertainties}. For the A-type regions, since the fit is performed in the \meff intervals, the normalisation is applied to all bins equally. (b) Results of the background-only fit extrapolated to VRs for the A-, B- and C-type regions. The normalisation of the backgrounds is obtained from the fit to the CRs. The upper panel shows the observed number of events and the predicted background yields. Statistical and systematic uncertainties as introduced in Section~\ref{sec:systematic_uncertainties} are included in the uncertainty band. The lower panel shows the significance in each VR. The significance calculation is performed as described in Ref.~\cite{SignificanceCalc}. The minor backgrounds (diboson, $W$+jets and \ttbar$+W/Z/h$) are grouped and labelled as `Other'.}
\label{fig:CR_Yields_Scales}
\end{figure}

\begin{landscape}
\begin{table}[h!]
\caption{Summary of all control and validation region definitions used in the analysis.}\label{tab:summaryCRVR}
\centering
\begin{tabular}{p{4cm}p{1cm}|cccc|cccc}
& & \multicolumn{4}{c}{Control Regions} & \multicolumn{4}{c}{Validation Regions} \\
\toprule
Variable       & Units & \CRA      & \CRB      & \CRCTop   & \CRCZ     & \VRA      & \VRB      & \VRCTop   & \VRCZ \\
\midrule
\met Trigger   &       & \ding{51} & \ding{51} & \ding{51} & -         & \ding{51} & \ding{51} & \ding{51} & \ding{51} \\
Lepton Trigger &       & -         & -         & -         & \ding{51} & -         & -         & -         & -     \\
\met           & [\GeV] & $>250$    & $>300$    & $>250$    & $<70$     & $>350$    & $>350$    & $>250$    & $\in[250,600]$\\
\dphimin       & [rad] & -         & -         & -         & $>0.2$    & $>0.4$    & $>0.4$    & $\in[0.2,0.4]$ & $>1.2$ \\
\midrule
\nlep(baseline) &       &= 1        &= 1        & = 1        &= 2             & = 0         & = 0         & = 0      & = 0 \\
\nlep(signal)   &       &= 1        &= 1        & = 1        &= 2(SFOS)       & -         & -         & -      & - \\
$\pt(\ell_1)$   & [\GeV] & $>20$     & $>20$     &  $>20$     & $>27$          & -         & -         & -      & - \\
$\pt(\ell_2)$   & [\GeV] & -         & -         &  -         & $>20$          & -         & -         & -      & -  \\
\mt             & [\GeV] & $>20$     & $>20$     &  $>20$     & -              & -         & -         & -      & -  \\
\mll            & [\GeV] & -         & -         &  -         & $\in[86,106]$  & -         & -         & -      & -  \\
$\tau$ veto     &       & -         & \ding{51} &  -         & -              & \ding{51} & \ding{51} & -      & -  \\
\midrule
\njets                   &       & $\geq 6$   &  $\geq 4$  & $\geq 4$  & $\geq 4$ & $\geq 6$   & $\geq 4$  & $\geq 4$ & $\geq 4$ \\
\nbjets                  &       & $\geq 4$   &  $\geq 4$  & $\geq 3$  & $\geq 3$ & $= 3$        & $= 3$       & $\geq 3$ & $= 2$       \\
$p_{\mathrm{T}}(b_1)$    & [\GeV] & $>100$ &  -         & -         & -        & $>100$ & -         & -        & -       \\
$p_{\mathrm{T}}(j_1)$    & [\GeV] & -          &  $> 350$   & -         & -        & -          & $>350$    & -       & -        \\
Leading jet not b-tagged &       & -          &  \ding{51} & -         & -        & -          & \ding{51} & -          & -        \\
$|\Delta\phi(j_1,\met)|$ & [rad] & -          &  $> 2.2$   & -         & -        & -          & $>2.8$    & -          & -        \\
\midrule
\METprime   & [\GeV] & -      &  -       & -     & $>250$ & -      & -      & -              & -            \\
\metsigObj  &       & -      &  -       & $>17$ & -      & $< 22$ & $<22$  & $\in[15,22]$ & $\in[20,22]$ \\
\meff       & [\TeV] & $>1.0$ &  $> 1.0$ & -     & -      & $>1.0$ & $>1.0$ & -            & -              \\
\bottomrule
\end{tabular}
\end{table}
\end{landscape}

 
\section{Systematic uncertainties}
\label{sec:systematic_uncertainties}
Several sources of experimental and theoretical systematic uncertainty on the signal and background
estimates are considered in this analysis. Their impact is reduced by fitting the event yields and normalising the dominant
backgrounds in the CRs defined with kinematic selections resembling those of the corresponding SRs (see Section \ref{sec:background_estimation}).
Uncertainties due to the numbers of events in the CRs are also introduced in the fit for each region.
The magnitude of the contributions arising from detector, theoretical modelling and statistical uncertainties are summarized in Table~\ref{tab:systematics-all2}.
 
Dominant detector-related systematic uncertainties arise from the $b$-tagging efficiency and mis-tagging rates, and from the jet energy scale and resolution. In SRA and SRB, the contributions of these uncertainties are almost equivalent. In SRC, the $b$-tagging uncertainty is dominant. The systematic uncertainty on the $b$-tagging efficiency ranges from 4.5\% for \bjets with \pt $\in [35, 40]$\,GeV up to 7.5\% for \bjets with high \pt ($> 100$\,GeV). The $b$-tagging uncertainty is estimated by varying the $\eta$-, \pt- and flavour-dependent scale factors applied to each jet in the simulation within a range that reflects the systematic uncertainty in the measured tagging efficiency and mis-tag rates in data ~\cite{Aad:2019aic}. The uncertainties in the jet energy scale and resolution are based on their respective measurements in data~\cite{PERF-2016-04, PERF-2011-04}.
 
The uncertainties associated with lepton reconstruction and energy measurements have a negligible impact on the final results; however, the lepton, photon and jet-related uncertainties are propagated to the calculation of the \etmiss, and additional uncertainties due to the energy scale and resolution
of the soft term are included in the \etmiss.
 
The systematic uncertainties related to the modelling of the energy of jets and leptons in the simulation are propagated to \metsigObj. No additional uncertainty on the energy resolution is applied, as the resolutions are taken to be the maximum of the parameterised data and simulation resolutions when performing the calculation for both data and MC simulation.
 
Uncertainties in the modelling of the SM background processes from MC simulation and their theoretical cross-section uncertainties are also taken into account. The dominant uncertainties in SRA and SRB arise from theoretical and modelling uncertainties of the \ttbar background. They are computed as the difference between the predictions from nominal samples and those from additional samples differing in hard-scattering generator and parameter settings, or by using internal weights assigned to the events depending on the choice of renormalisation and factorisation scales, initial- and final-state radiation parameters, and PDF sets. The impact of the PS and hadronisation model is evaluated by comparing the nominal generator with a \powheg\ sample interfaced to \herwig7~\cite{Bahr:2008pv,Bellm:2015jjp}, using the H7UE set of tuned parameters~\cite{Bellm:2015jjp}. To assess the uncertainty due to the choice of hard-scattering generator and matching scheme, an alternative generator setup using \amcpeight is employed. It uses the shower starting scale, $\mu_{\mathrm q} = H_{\mathrm T}/2$, where $H_{\mathrm T}$ is defined here as the scalar sum of the \pT of all outgoing partons.
 
The dominant uncertainties in SRC arise from the MC modelling of the $Z$+jets process, followed by the $\ttbar$ and single-top modelling. The \Zjets\ (as well as \Wjets) modelling uncertainties are estimated by considering different merging (CKKW-L) and resummation scales using alternative samples, PDF variations from the NNPDF3.0 NNLO replicas~\cite{Gleisberg:2008ta}, and variations of factorisation and renormalisation scales in the ME. The latter have been evaluated using 7 point-variations, changing the renormalisation and factorisation scales up and down by factors 0.5 and 2, such that when one scale is up the other is down, and vice-versa.
 
For the SUSY signal processes, both the experimental and theoretical uncertainties in the expected signal yield are considered.
Experimental uncertainties are found to be 6--36\% across the mass plane with fixed LSP mass for A-type SRs, and 4--40\% for C-type SRs. For models where \dmNN~=~130 GeV is assumed, scenarios where SRB is relevant have uncertainties of 11--37\%.
 
In all SRs, the dominant uncertainty on the signal yields is found to be from the $b$-tagging efficiency.
 
Theoretical uncertainties in the approximate NNLO+NNLL cross-section are calculated for each
SUSY signal scenario, and are dominated by the uncertainties in the renormalisation and factorisation scales, followed
by the uncertainty in the PDFs. These are 7--17\% for bottom-squark masses in the range between 400~\GeV\ and 1500~\GeV.
Additional uncertainties in the acceptance and efficiency due to the modelling of ISR and CKKW scale variations
in SUSY signal MC samples are also taken into account, and contribute up to $\sim$10\%.

\begin{table}[h!]
\footnotesize{
\begin{center}
\caption{Expected background event yields and dominant systematic uncertainties on background estimates in the A-type (inclusive), B-type and C-type (inclusive) regions.
Individual uncertainties can be correlated, and do not necessarily add up quadratically to
the total background uncertainty. The percentages show the size of the uncertainty relative to the total expected background.}\label{tab:systematics-all2}
\setlength{\tabcolsep}{0pt}
\begin{tabular*}{\textwidth}{@{\extracolsep{\fill}}lrrrrrrrrr}
\toprule
Region                                  & \multicolumn{2}{c}{SRA} & & \multicolumn{2}{c}{SRB} & & \multicolumn{2}{c}{SRC} &        \\
\midrule
Total background expectation                           &  \multicolumn{2}{c}{17.1} & & \multicolumn{2}{c}{3.3} &  & \multicolumn{2}{c}{37.9} &  \\
\midrule
Total background uncertainty                                 & $ 2.8$ & (16\%) & & $ 0.9$  &  (27\%) & & $ 6.2$  &  (16\%) & \\
\midrule
Systematic, experimental                               & $ 1.4$  &  (8\%) & & $ 0.3$  &  (10\%) & & $ 3.0$ & (8\%) &\\
Systematic, theoretical                                & $ 2.3$  &  (13\%) & & $ 0.6$  &  (18\%) & & $ 3.2$& (8\%) & \\
Statistical, MC samples                                & $ 0.7$  &  (4\%) & & $ 0.4$ & (12\%) & & $ 2.0$ & (5\%) & \\
\bottomrule
\end{tabular*}
\end{center}
}
\end{table}
 
\section{Results and interpretation}
\label{sec:results_and_interpretation}
The event yields for all SRs are reported in Table~\ref{tab:fit_results}.
The SM background expectations resulting from background-only fits are also reported
showing statistical plus systematic uncertainties.
The largest background contribution in A-type and B-type SRs arises from \ttbar~ production, whilst the contribution from $Z\rightarrow \nu\bar{\nu}$ production in association with $b$-quarks is largest in the C-type SRs, with sub-dominant contributions from the \ttbar and single-top processes. Other background sources are \ttbar+$W/Z$, \ttbar+$h$, diboson and $W$+jets production.
The results are also summarised in Figure~\ref{fig:SRpulls}, where the significances for each of the SRs are also presented.
No significant deviations are observed between expected and observed yields in all signal regions considered.
 
\begin{figure}[!h]
\begin{center}
\includegraphics[width=0.75\textwidth]{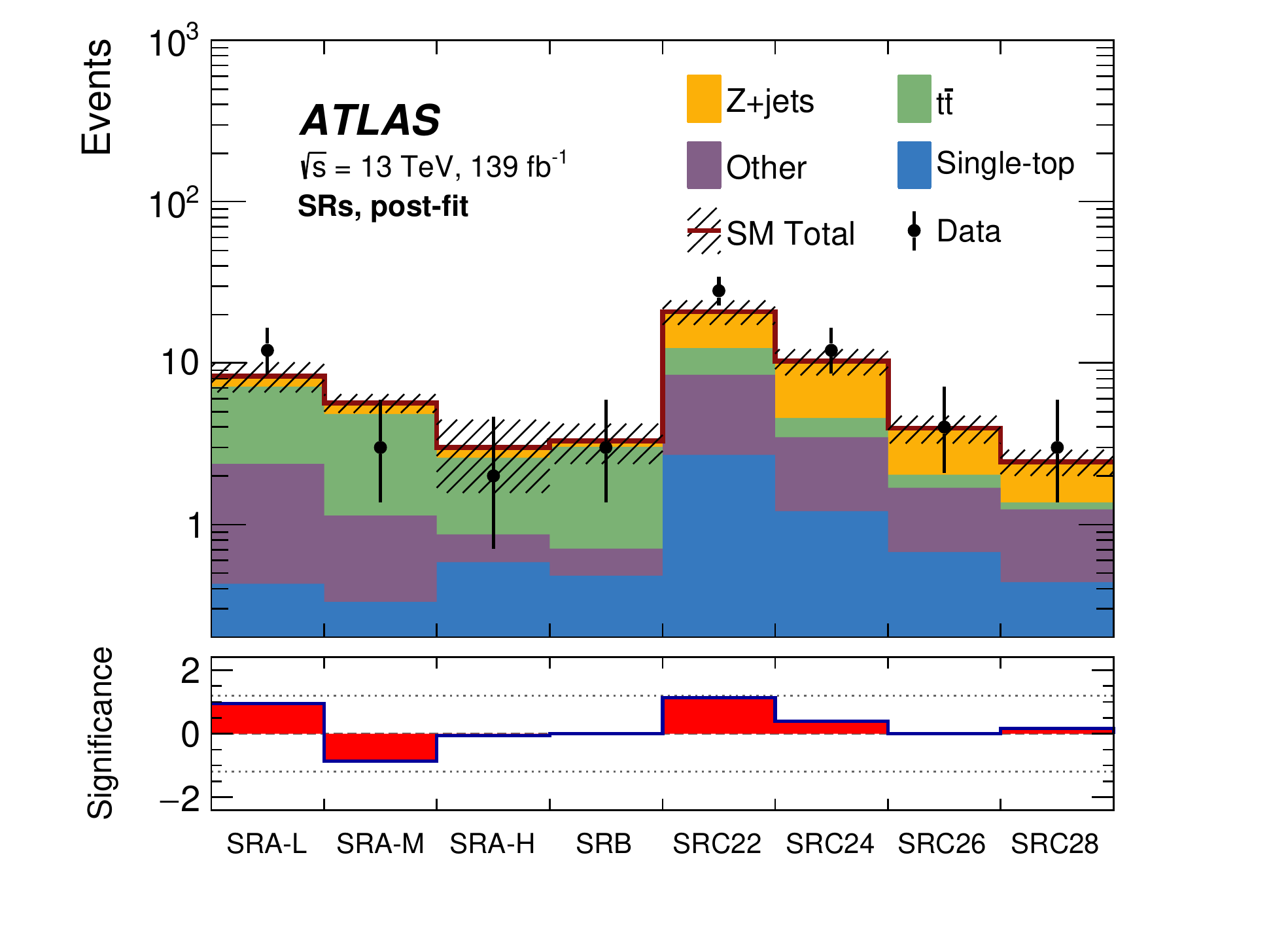}
\end{center}
\caption{Results of the background-only fit extrapolated to all SRs.
The normalisation of the backgrounds is obtained from the fit to the CRs.
The upper panel shows the observed number of events and the predicted background yields. The backgrounds which contribute only a small amount (diboson, $W$+jets and \ttbar$+W/Z/h$) are grouped and labelled as ``Other''.
All uncertainties defined in Section~\ref{sec:systematic_uncertainties}
are included in the uncertainty band.  The lower panel shows the significance in each SR. The significance calculation is performed as described in Ref.~\cite{SignificanceCalc}. }
\label{fig:SRpulls}
\end{figure}
 
Figure~\ref{fig:SR_N1s} shows comparisons between the observed data
and the post-fit SM predictions for some relevant kinematic distributions for the inclusive SRA, SRB and SRC
selections before selection requirements are applied on the quantity shown. The expected distributions for scenarios with different bottom squark, \neuttwo and \neut masses (depending on the SR considered) are shown for illustrative purposes.
 
\sisetup{parse-numbers=false,input-uncertainty-signs=true}
\begin{table}[]
\begin{center}
\caption{Background-only fit results for the A- and B-type regions (top table) and C-type regions (bottom table) performed using 139\,\ifb of data. The quoted uncertainties on the fitted SM background include both the statistical and systematic uncertainties.}
\label{tab:fit_results}
\end{center}
\setlength{\tabcolsep}{0pt}
\resizebox{\columnwidth}{!}{
{\small
\sisetup{retain-explicit-plus=true}
\begin{tabular*}{\textwidth}{@{\extracolsep{\fill}}l rcl c rcl c rcl c rcl c rcl}
\noalign{\smallskip}\hline\noalign{\smallskip}
{ } & \multicolumn{3}{c}{SRA} &  & \multicolumn{3}{c}{SRA-L} & & \multicolumn{3}{c}{SRA-M} &  & \multicolumn{3}{c}{SRA-H} &  & \multicolumn{3}{c}{SRB} \\[-0.05cm]
\noalign{\smallskip}\hline\noalign{\smallskip}
Observed events & \multicolumn{3}{c}{$17$}    &   & \multicolumn{3}{c}{$12$}  &    & \multicolumn{3}{c}{$3$} &  & \multicolumn{3}{c}{$2$}  &  & \multicolumn{3}{c}{$3$}          \\
\noalign{\smallskip}\hline\noalign{\smallskip}
Fitted SM bkg events                                      & $17.1$&$\pm$&$2.8$               &     &$8.4$&$\pm$&$1.7$                &       &$5.7$&$\pm$&$0.8$      &    &$3.0$&$\pm$&$1.5$ &  &$3.3$&$\pm$&$0.9$  \\
\noalign{\smallskip}\hline\noalign{\smallskip}
\ttbar                                              & $10.1$ & $\pm$ & $2.5$ & & $4.7$ & $\pm$ & $1.5$ & & $3.7$ & $\pm$ & $0.6$ & & $1.7$ & $\pm$ & $1.4$ & & $2.3$  & $\pm$ & $0.8$\\
\Zjets                                              & $2.6$  & $\pm$ & $0.4$ & & $1.3$ & $\pm$ & $0.2$ & & $0.9$ & $\pm$ & $0.2$ & & $0.4$ & $\pm$ & $0.1$ & & $0.3$  & $\pm$ & $0.1$ \\
Single-top                                          & $1.4$  & $\pm$ & $0.3$ & & $0.4$ & $\pm$ & $0.1$ & & $0.3$ & $\pm$ & $0.1$ & & $0.6$ & $\pm$ & $0.2$ & & $0.5$  & $\pm$ & $0.1$ \\
\ttV                                                & $1.2$  & $\pm$ & $0.3$ & & $0.7$ & $\pm$ & $0.1$ & & $0.3$ & $\pm$ & $0.1$ & & $0.1$ & $\pm$ & $0.0$ & & $0.07$ & $\pm$ & $0.02$   \\
\ttbar+$h$                                          & $1.1$  & $\pm$ & $0.2$ & & $0.7$ & $\pm$ & $0.1$ & & $0.3$ & $\pm$ & $0.1$ & & $0.1$ & $\pm$ & $0.0$ & & $0.13$ & $\pm$ & $0.02$ \\
\Wjets                                              & $0.4$  & $\pm$ & $0.1$ & & $0.2$ & $\pm$ & $0.1$ & & $0.1$ & $\pm$ & $0.0$ & &       & {--}  &       & & $0.02$ & $\pm$ & $0.01$       \\
Diboson                                             & $0.4$  & $\pm$ & $0.1$ & & $0.3$ & $\pm$ & $0.1$ & &       & {--}  &       & &       & {--}  &       & &        & {--}  & \\
\noalign{\smallskip}\hline\noalign{\smallskip}
$m(\sbot, \neuttwo, \neut) = (1100, 330, 200)$\,GeV & $13.7$ & $\pm$ & $0.3$ & & $0.7$ & $\pm$ & $0.1$ & & $6.3$ & $\pm$ & $0.2$ & & $6.6$ & $\pm$ & $0.2$ & & $0.3$  & $\pm$ & $0.1$  \\
$m(\sbot, \neuttwo, \neut) = (700, 680, 550)$\,GeV  & $1.3$  & $\pm$ & $0.6$ & & $0.2$ & $\pm$ & $0.1$ & & $0.5$ & $\pm$ & $0.4$ & & $0.6$ & $\pm$ & $0.4$ & & $7.4$  & $\pm$ & $1.2$  \\
$m(\sbot, \neuttwo, \neut) = (1200, 1150, 60)$\,GeV & $8.7$  & $\pm$ & $0.2$ & & $1.4$ & $\pm$ & $0.1$ & & $3.4$ & $\pm$ & $0.1$ & & $3.8$ & $\pm$ & $0.1$ & & $0.6$  & $\pm$ & $0.1$  \\
\noalign{\smallskip}\hline\noalign{\smallskip}
\end{tabular*}
}
}
{\small
\sisetup{retain-explicit-plus=true}
\begin{tabular*}{\textwidth}{@{\extracolsep{\fill}}l rcl c rcl c rcl c rcl c rcl}
\noalign{\smallskip}\hline\noalign{\smallskip}
{}  & \multicolumn{3}{c}{SRC}   &    & \multicolumn{3}{c}{SRC22}   &   & \multicolumn{3}{c}{SRC24}  &	& \multicolumn{3}{c}{SRC26} &  & \multicolumn{3}{c}{SRC28}      \\[-0.05cm]
\noalign{\smallskip}\hline\noalign{\smallskip}
Observed events          &  \multicolumn{3}{c}{$47$}        &     &  \multicolumn{3}{c}{$28$}    &          &  \multicolumn{3}{c}{$12$}              &  & \multicolumn{3}{c}{$4$}             &   & \multicolumn{3}{c}{$3$}                   \\
\noalign{\smallskip}\hline\noalign{\smallskip}
Fitted SM bkg events         &  $37.9$ & $\pm$ & $6.2$         & &  $21.2$ & $\pm$ & $4.1$   &       &  $10.6$ & $\pm$ & $2.3$    &      &  $3.7$ & $\pm$ & $0.9$     &     &  $2.4$ & $\pm$ & $0.6$              \\
\noalign{\smallskip}\hline\noalign{\smallskip}
\ttbar          &  $5.4$ & $\pm$ & $2.6$   &       &  $3.9$ & $\pm$ & $2.3$    &      &  $1.1$ & $\pm$ & $0.6$     &     &  $0.3$ & $\pm$ & $0.3$    &      &  $0.1$ & $\pm$ & $0.1$              \\
\Zjets          &  $17.6$ & $\pm$ & $4.7$      &    &  $8.8$ & $\pm$ & $2.5$     &     &  $6.0$ & $\pm$ & $1.8$      &    &  $1.7$ & $\pm$ & $0.7$     &     &  $1.1$ & $\pm$ & $0.4$              \\
Single-top          &  $5.0$ & $\pm$ & $1.5$    &      &  $2.7$ & $\pm$ & $1.0$       &   &  $1.2$ & $\pm$ & $0.3$   &       &  $0.7$ & $\pm$ & $0.2$       &   &  $0.4$ & $\pm$ & $0.1$              \\
\ttV          &  $4.3$ & $\pm$ & $0.6$       &   &  $2.5$ & $\pm$ & $0.4$      &    &  $1.1$ & $\pm$ & $0.2$    &      &  $0.5$ & $\pm$ & $0.1$   &       &  $0.2$ & $\pm$ & $0.1$              \\
\ttbar+$h$          &  $0.2$ & $\pm$ & $0.0$    &      &  $0.2$ & $\pm$ & $0.0$      &    &   & -- &     &    &  $0.1$ & $\pm$ & $0.0$    &      &  $0.0$ & $\pm$ & $0.0$              \\
\Wjets          &  $3.5$ & $\pm$ & $0.8$     &     &  $2.2$ & $\pm$ & $0.5$     &     &  $0.6$ & $\pm$ & $0.2$       &   &  $0.2$ & $\pm$ & $0.1$      &    &  $0.4$ & $\pm$ & $0.1$              \\
Diboson          &  $1.8$ & $\pm$ & $0.3$    &      &  $0.9$ & $\pm$ & $0.2$    &      &  $0.6$ & $\pm$ & $0.1$    &      &  $0.2$ & $\pm$ & $0.0$    &      &  $0.1$ & $\pm$ & $0.1$              \\
\noalign{\smallskip}\hline\noalign{\smallskip}
$m(\sbot, \neuttwo, \neut) = (1100, 330, 200)$\,GeV                    &  $0.4$ & $\pm$ & $0.1$    &     &  $ 0.3$ & $\pm$ & $0.1$       &          &  $ 0.1$ & $\pm$ & $0.0$       &         &  $0.03$ & $\pm$ & $0.02$ &   &  $ 0.03$ & $\pm$ & $0.01$   \\
$m(\sbot, \neuttwo, \neut) = (700, 680, 550)$\,GeV                    &  $1.2$ & $\pm$ & $0.5$    &     &  $ 0.5$ & $\pm$ & $0.2$       &          &  $ 0.7$ & $\pm$ & $0.4$        &        &  & -- & &  &  & -- &  \\
$m(\sbot, \neuttwo, \neut) = (1200, 1150, 60)$\,GeV                    &  $26.7$ & $\pm$ & $0.3$  &       &  $ 6.3$ & $\pm$ & $0.2 $     &            &  $ 6.4$ & $\pm$ & $0.2 $        &        &  $ 5.8$ & $\pm$ & $0.2 $  &   &  $ 8.3$ & $\pm$ & $0.2 $   \\
\noalign{\smallskip}\hline\noalign{\smallskip}
 
\end{tabular*}
}
\end{table}

\begin{figure}[!h]
\begin{center}
\subfloat[]{\includegraphics[width=0.45\textwidth]{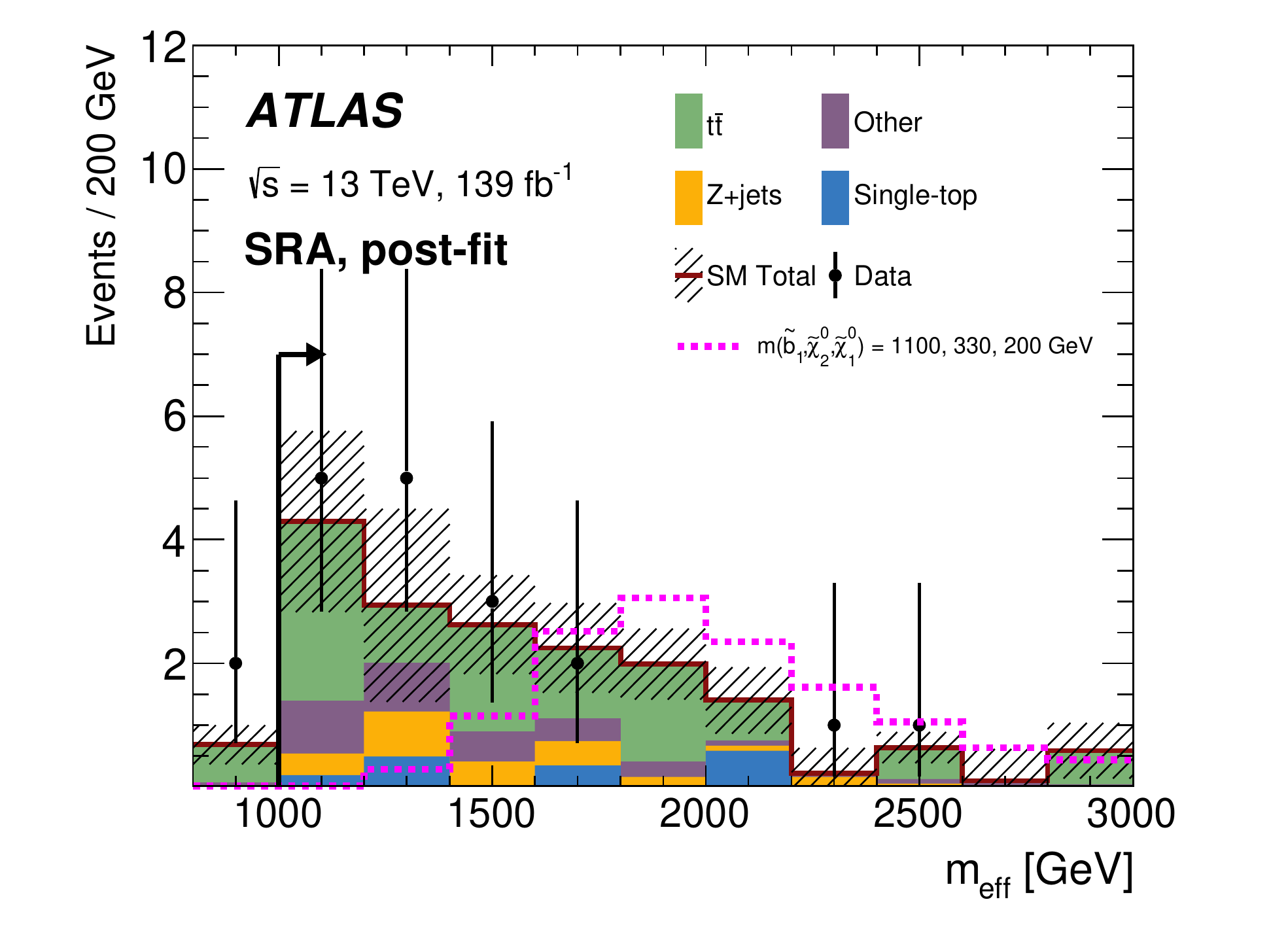}}
\subfloat[]{\includegraphics[width=0.45\textwidth]{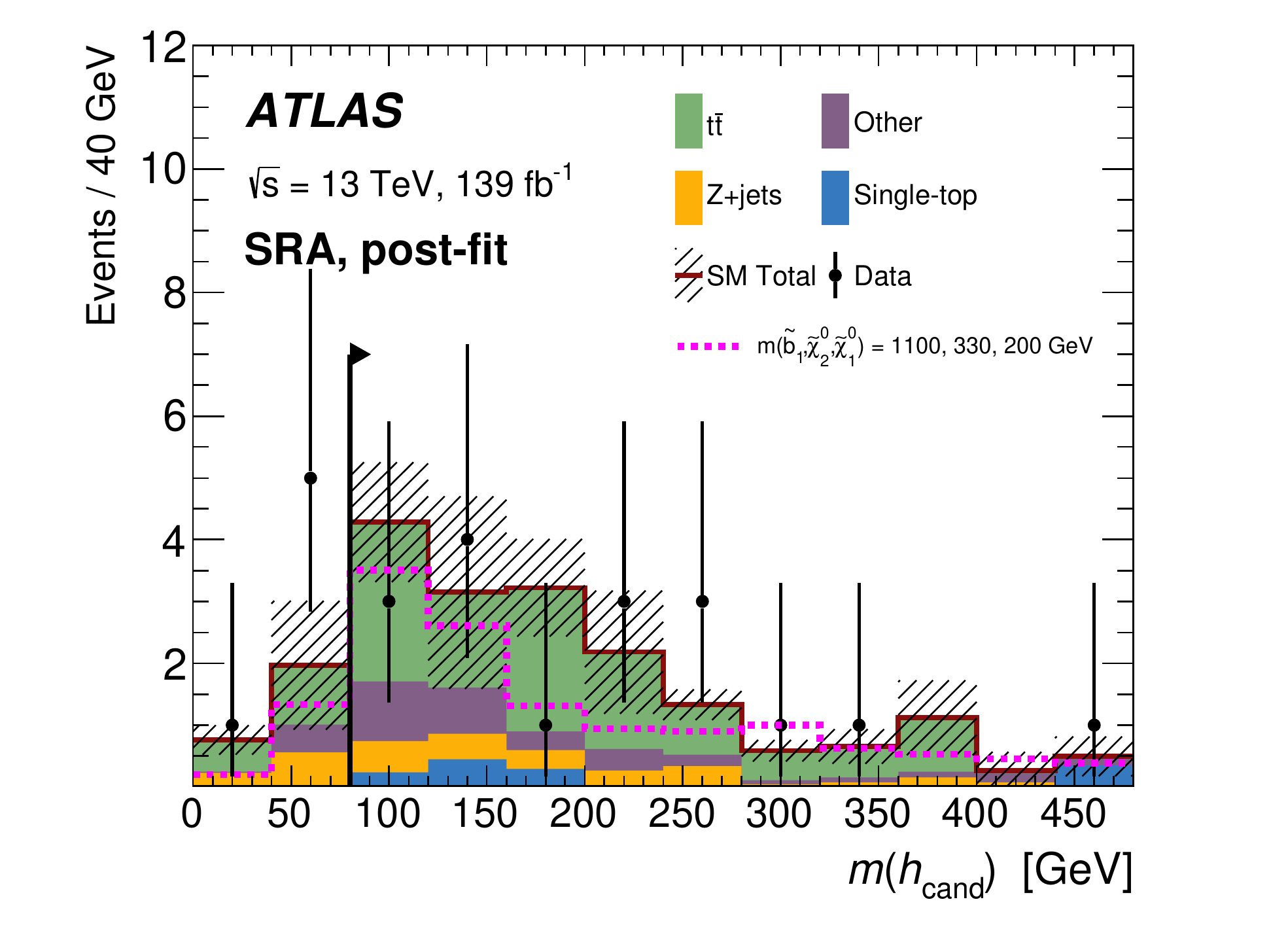}}\\
\subfloat[]{\includegraphics[width=0.45\textwidth]{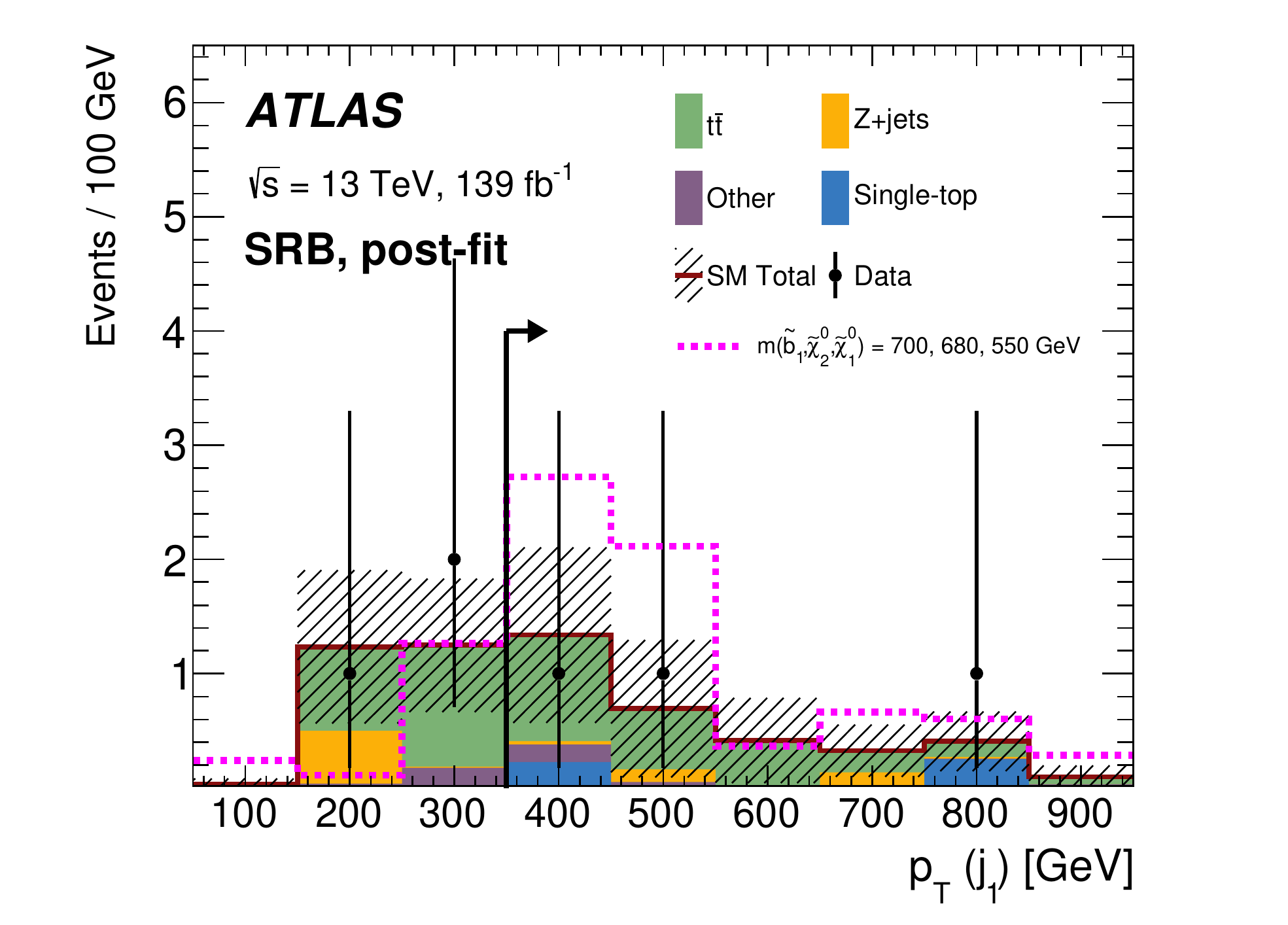}}
\subfloat[]{\includegraphics[width=0.45\textwidth]{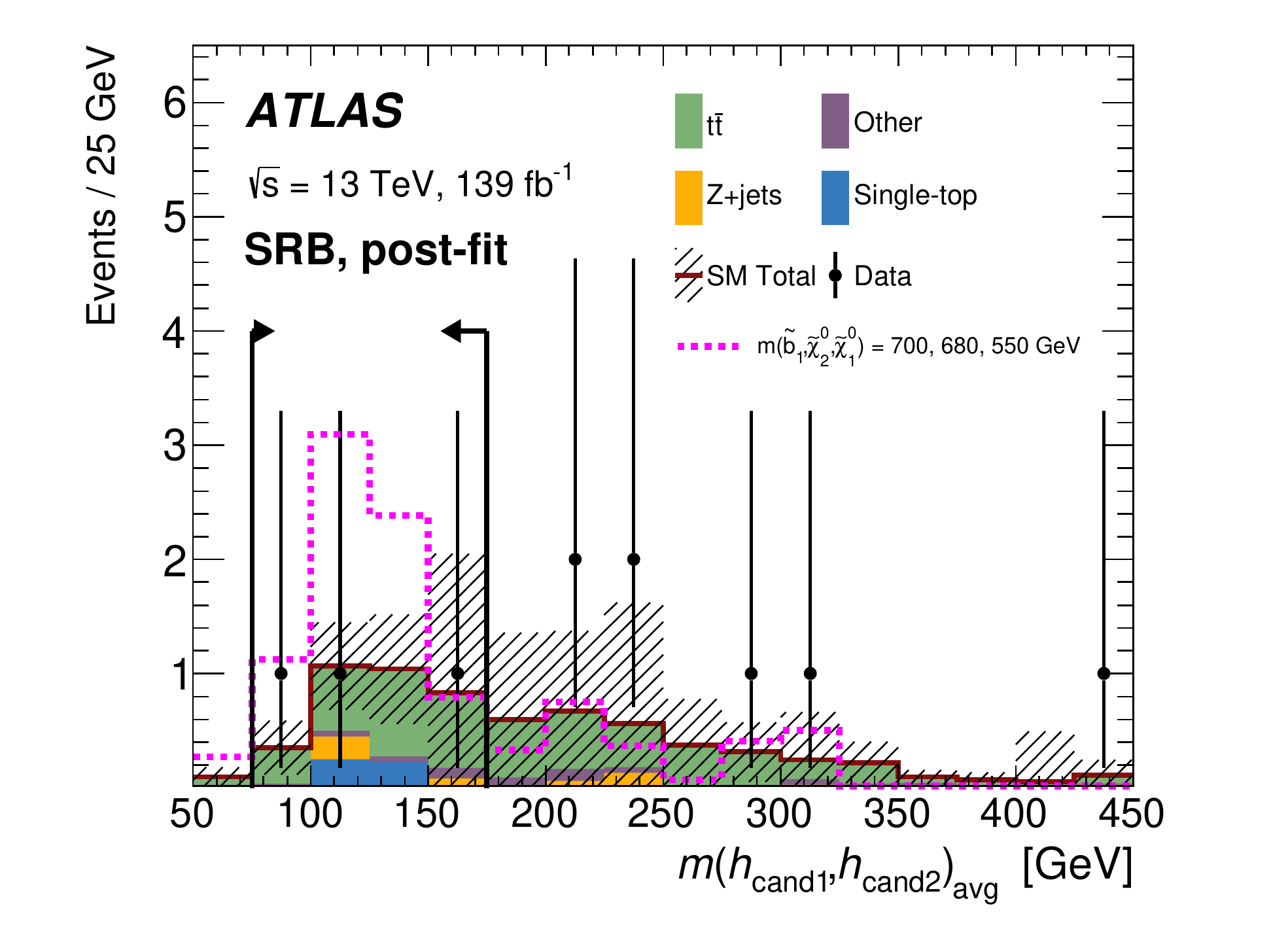}}\\
\subfloat[]{\includegraphics[width=0.45\textwidth]{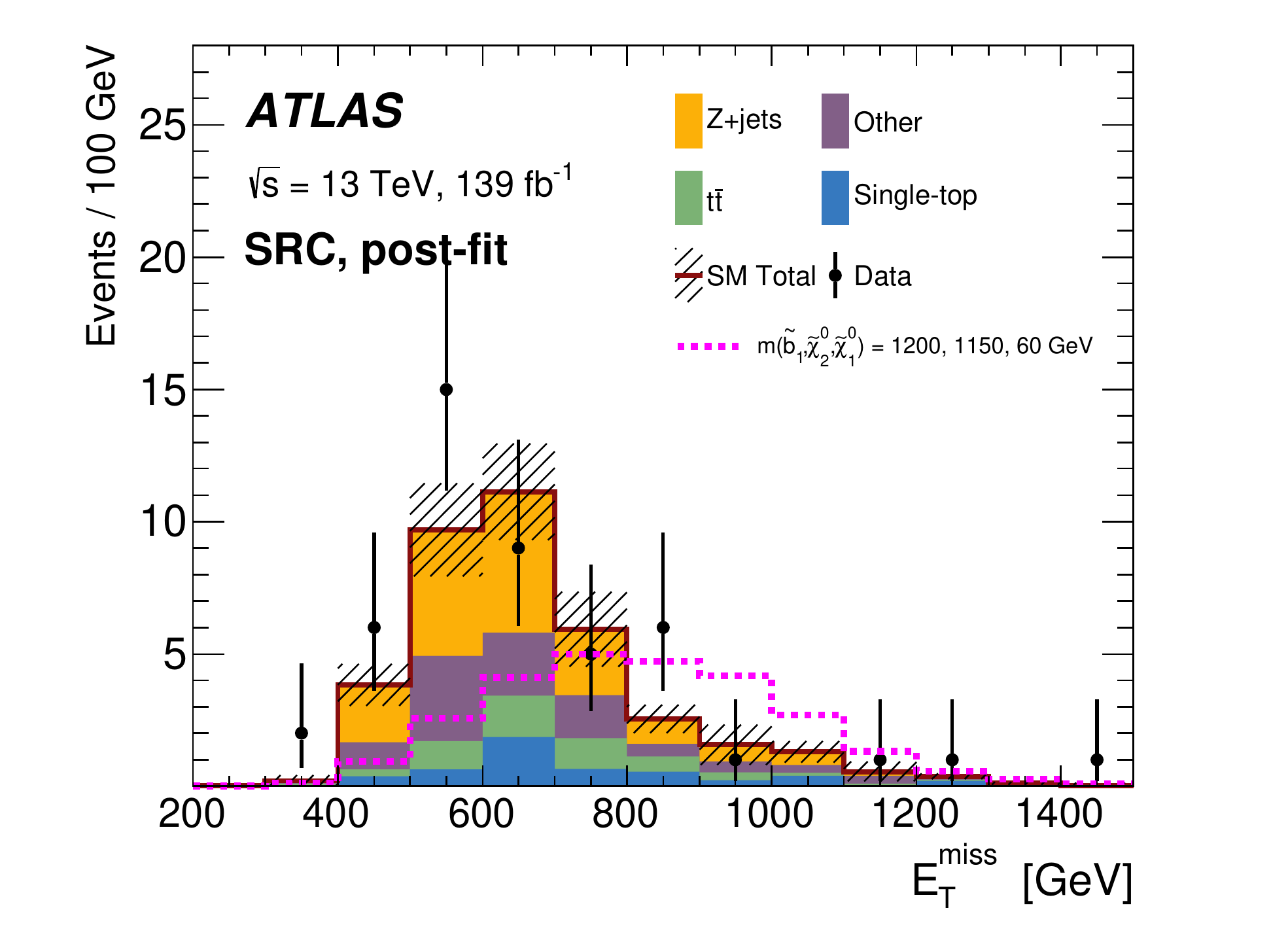}}
\subfloat[]{\includegraphics[width=0.45\textwidth]{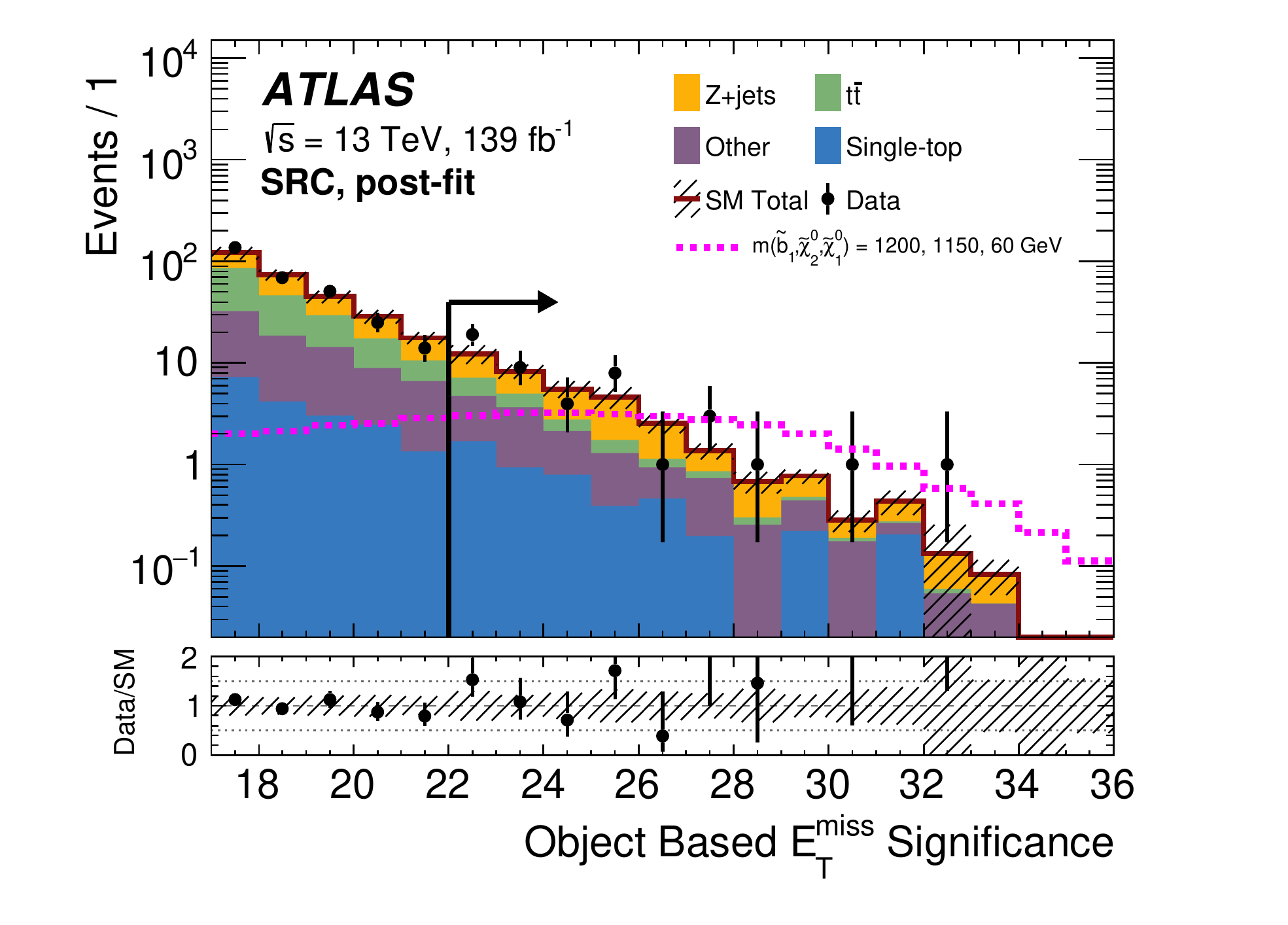}}
\end{center}
\caption{Post-fit distributions of (a) the \meff and (b) the \maxminDRmbb in the inclusive SRA region; (c) the leading jet \pt and (d) the $m(h_{\mathrm{cand1}},h_{\mathrm{cand2}})_{\mathrm{avg}}$ for the Higgs candidates in the SRB region; (e) \met and (f) \metsigObj for SRC regions. All SR selections are applied except for the selection on the variable shown, where the selection on the variable under consideration is denoted by an arrow, except in the case of (e), where the full SRC selection is applied. All uncertainties as defined in Section~\ref{sec:systematic_uncertainties} are included in the uncertainty band. The minor backgrounds (diboson, $W$+jets and \ttbar$+W/Z/h$) are grouped and labelled as ``Other''. For illustration, contributions expected for scenarios with different bottom-squark, \neuttwo and \neut masses depending on the SR considered are superimposed. Overflow events which do not fall into the axis range are placed into the right-most bin.}
\label{fig:SR_N1s}
\end{figure}
 
The CL$_{\mathrm{s}}$ technique \cite{Read:2002hq} is used to place 95\% Confidence Level (CL) upper limits on event yields from physics beyond the SM (BSM) for each signal region.
The profile-likelihood-ratio test statistic is used to exclude the signal-plus-background hypothesis for
specific signal models. When normalised to the integrated luminosity of the data sample, results can be interpreted
as corresponding upper limits on the visible cross-section, $\sigma^{}_\textrm{vis}$, defined as the product of
the BSM production cross-section, the acceptance and the selection efficiency of a BSM signal. When calculating the model-independent upper limits of the A- and C-type regions, only the inclusive SR selection is used.
\Tab{\ref{tab:upper}} summarises the observed ($S^{95}_\mathrm{obs}$)
and expected ($S^{95}_\mathrm{exp}$) 95\% CL upper limits on the number of BSM events and on $\sigma^{}_\textrm{vis}$
for all SRs. The $p_0$-values, which represent the probability of the SM background to
fluctuate to the observed number of events or higher, are also provided and are capped at $p_0 = 0.5$; the associated significance is provided in parentheses.
 
\begin{table}[h!]
\footnotesize{
\begin{center}\renewcommand\arraystretch{1.2}
\caption{From left to right, observed 95\% CL upper limits on the visible cross sections $\sigma^{}_\textrm{vis}$,
the observed ($S^{95}_\mathrm{obs}$) and expected ($S^{95}_\mathrm{exp}$) 95\% CL upper limits on the number of
signal events with $\pm$ 1 $\sigma$ excursions of the expectation, the CL of the background-only hypothesis, $\mathrm{CL}_{\mathrm{B}}$, the discovery $p$-value ($p_0$),
truncated at 0.5, and the associated significance (in parentheses). }
\label{tab:upper}
\begin{tabular*}{\textwidth}{@{\extracolsep{\fill}}lrrrrr}
\noalign{\smallskip}\hline\noalign{\smallskip}
{\textbf {Signal channel}}                        & $\sigma_{\mathrm {vis}}$[fb]  &  $S_{\mathrm {obs}}^{95}$  & $S_{\mathrm {exp}}^{95}$ & $\mathrm{CL}_{\mathrm{B}}$ & $p_0$ ($Z$)  \\
\noalign{\smallskip}\hline\noalign{\smallskip}
SRA    & $0.06$ &  $9.0$ & $ { 10.1 }^{ +4.7 }_{ -3.1 }$ & $0.38$ & $ 0.50$~$(0.00)$ \\%
SRB    & $0.04$ &  $4.9$ & $ { 5.1 }^{ +2.8 }_{ -1.7 }$ & $0.45$ & $ 0.50$~$(0.00)$ \\%
SRC    & $0.19$ &  $26.0$ & $ { 20.8 }^{ +7.0 }_{ -5.5 }$ & $0.80$ & $ 0.17$~$(0.97)$ \\%
\noalign{\smallskip}\hline\noalign{\smallskip}
\end{tabular*}
\end{center}
}
\end{table}

Model-dependent exclusion limits are obtained assuming the two types of SUSY particle mass hierarchies described in Section \ref{sec:intro}.
The lightest bottom squark decays exclusively via $\bone \rightarrow b \neuttwo$ with subsequent decay $\neuttwo \rightarrow h \neut$.
The expected limits from the SRs are compared for each set of scenarios and the observed limits are
obtained by choosing the SR with the best expected sensitivity for each SUSY model.
The fit procedure takes into account correlations in the yield predictions between
control and signal regions due to common background normalisation parameters and systematic uncertainties.
The experimental systematic uncertainties in the signal are taken into account for the calculation
and are assumed to be fully correlated with those in the SM background.
 
Figures~\ref{fig:M60_Best_Limits} and~\ref{fig:DM130_Best_Limits} show the observed
(solid line) and expected (dashed line) exclusion contours at 95\% CL in the $\bone$--$\neuttwo$
mass planes for the two types of SUSY scenarios considered.
For the scenarios where the mass of the neutralino is assumed to be 60 GeV, the sensitivity to models with
the largest mass difference between the $\bone$ and the $\neuttwo$~is achieved
with the combination of the A-type SRs. Sensitivity to scenarios with small
mass differences is obtained with the dedicated C-type SRs.
For scenarios with \dmNN~=~130~GeV, the sensitivity of the A-type SRs is complemented by the B-type SR in the case of small mass
difference between the $\bone$ and the $\neuttwo$.
 
\begin{figure}[!htb]
\centering
\subfloat[]{\includegraphics[width=0.76\columnwidth]{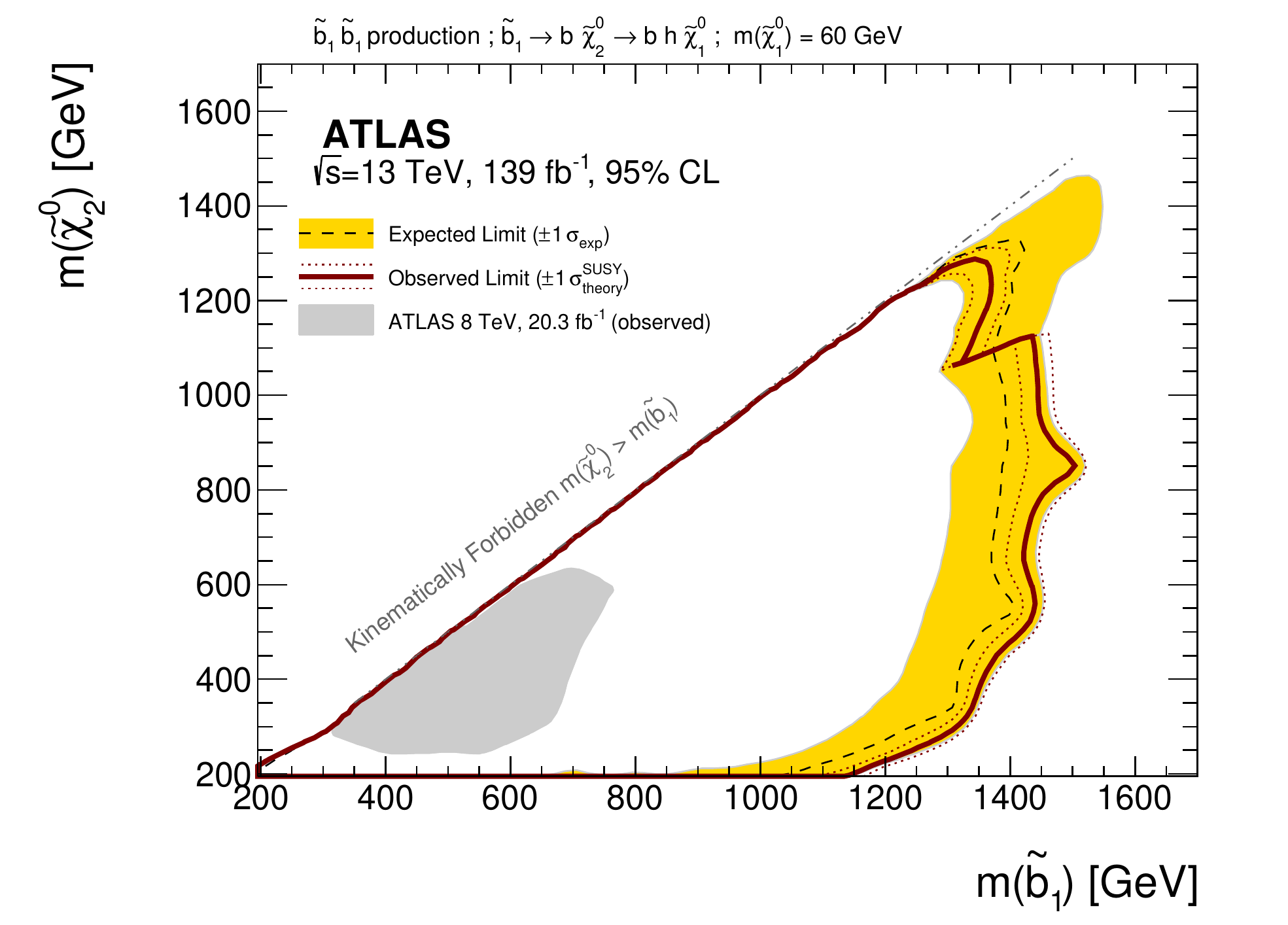}\label{fig:M60_Best_Limits}} \\
\subfloat[]{\includegraphics[width=0.76\columnwidth]{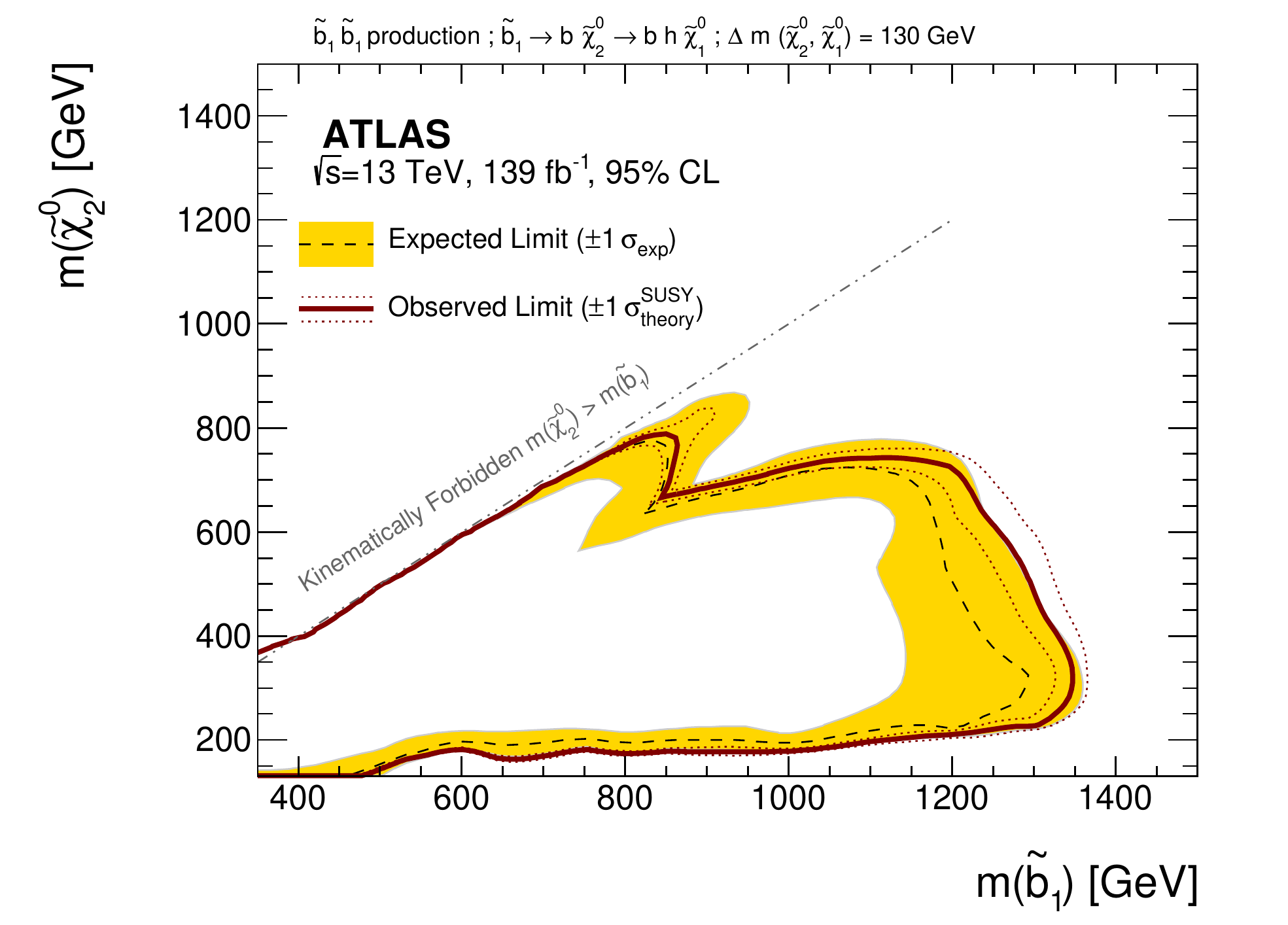}\label{fig:DM130_Best_Limits}}\\
\caption{Exclusion contour at the 95\% CL in the $m(\sbot,\neuttwo)$ phase space for (a) the $m(\neut) = 60$\,GeV signal scenario, ATLAS Run 1 limit taken from Ref.~\cite{Aad:2015pfx} and (b) the $\Delta m(\neuttwo, \neut) = 130$\,GeV signal scenario, using the SR with the best-expected sensitivity. The theory uncertainty band contains the systematic uncertainties on the signal model under consideration and the uncertainty in the signal cross section. }
\label{fig:Best_expectedLimits}
\end{figure}
 
Bottom-squark masses up to 1.5~TeV are excluded for models with fixed $m_{\neut} = 60$\,GeV and
\neuttwo masses $\in [0.5,1.1]$\,TeV. 
In case of \dmNN~=~130~GeV, bottom-squark masses up to 1.3~TeV are excluded for \neuttwo masses up to 750\,GeV.
The losses in sensitivity for models where \neuttwo~masses are below 190\,GeV are due to the stringent
requirements on \etmiss.
 
The results constitute a large improvement upon previous Run-1 searches and significantly strengthen the constraints on bottom squark
masses; they are also complementary to other searches where bottom squarks are assumed to decay directly
to a bottom-quark and a neutralino or to a top-quark and a chargino \cite{Aaboud:2017wqg}.
 
\FloatBarrier
 
\section{Conclusion}
\label{sec:conclusion}
The result of a search for pair production of bottom squarks is reported. The analysis uses $139$\,fb$^{-1}$ of $pp$
collisions at $\sqrt{s}=13$~\TeV\ collected by the ATLAS experiment at the LHC between 2015 and 2018.
$R$-parity-conserving SUSY scenarios where bottom squarks decay into a $b$-quark and the second-lightest neutralino,
$\sbot \rightarrow b + \neuttwo$, with \neuttwo subsequently decaying into a Higgs boson like the one in the SM and the lightest neutralino, are considered.
The search investigates final states containing large missing transverse momentum and three or more \bjets.
No significant excess of events above the expected Standard Model background is found and exclusion limits at the 95\% confidence
level are placed on the visible cross-section and on the mass of the bottom squark for various assumptions about
the mass hierarchy of the \bone, \neuttwo and \neut.
Bottom-squark masses up to 1.5 (1.3)~TeV are excluded for \neuttwo masses up to 1100 (750)\,GeV in models with fixed $m({\neut}) = 60$\,GeV (\dmNN = 130 GeV).
As the first search for such scenarios carried out by ATLAS in Run 2, these results are a significant improvement upon the previous Run-1 result, considerably tightening the constraints on bottom-squark production. 
 

\section*{Acknowledgements}
 
 
We thank CERN for the very successful operation of the LHC, as well as the
support staff from our institutions without whom ATLAS could not be
operated efficiently.
 
We acknowledge the support of ANPCyT, Argentina; YerPhI, Armenia; ARC, Australia; BMWFW and FWF, Austria; ANAS, Azerbaijan; SSTC, Belarus; CNPq and FAPESP, Brazil; NSERC, NRC and CFI, Canada; CERN; CONICYT, Chile; CAS, MOST and NSFC, China; COLCIENCIAS, Colombia; MSMT CR, MPO CR and VSC CR, Czech Republic; DNRF and DNSRC, Denmark; IN2P3-CNRS, CEA-DRF/IRFU, France; SRNSFG, Georgia; BMBF, HGF, and MPG, Germany; GSRT, Greece; RGC, Hong Kong SAR, China; ISF and Benoziyo Center, Israel; INFN, Italy; MEXT and JSPS, Japan; CNRST, Morocco; NWO, Netherlands; RCN, Norway; MNiSW and NCN, Poland; FCT, Portugal; MNE/IFA, Romania; MES of Russia and NRC KI, Russian Federation; JINR; MESTD, Serbia; MSSR, Slovakia; ARRS and MIZ\v{S}, Slovenia; DST/NRF, South Africa; MINECO, Spain; SRC and Wallenberg Foundation, Sweden; SERI, SNSF and Cantons of Bern and Geneva, Switzerland; MOST, Taiwan; TAEK, Turkey; STFC, United Kingdom; DOE and NSF, United States of America. In addition, individual groups and members have received support from BCKDF, CANARIE, CRC and Compute Canada, Canada; COST, ERC, ERDF, Horizon 2020, and Marie Sk{\l}odowska-Curie Actions, European Union; Investissements d' Avenir Labex and Idex, ANR, France; DFG and AvH Foundation, Germany; Herakleitos, Thales and Aristeia programmes co-financed by EU-ESF and the Greek NSRF, Greece; BSF-NSF and GIF, Israel; CERCA Programme Generalitat de Catalunya, Spain; The Royal Society and Leverhulme Trust, United Kingdom.
 
The crucial computing support from all WLCG partners is acknowledged gratefully, in particular from CERN, the ATLAS Tier-1 facilities at TRIUMF (Canada), NDGF (Denmark, Norway, Sweden), CC-IN2P3 (France), KIT/GridKA (Germany), INFN-CNAF (Italy), NL-T1 (Netherlands), PIC (Spain), ASGC (Taiwan), RAL (UK) and BNL (USA), the Tier-2 facilities worldwide and large non-WLCG resource providers. Major contributors of computing resources are listed in Ref.~\cite{ATL-GEN-PUB-2016-002}.
 

\clearpage
\clearpage
 
\clearpage
 
\printbibliography
 
\clearpage
 
\begin{flushleft}
{\Large The ATLAS Collaboration}

\bigskip

G.~Aad$^\textrm{\scriptsize 101}$,    
B.~Abbott$^\textrm{\scriptsize 128}$,    
D.C.~Abbott$^\textrm{\scriptsize 102}$,    
O.~Abdinov$^\textrm{\scriptsize 13,*}$,    
A.~Abed~Abud$^\textrm{\scriptsize 70a,70b}$,    
K.~Abeling$^\textrm{\scriptsize 53}$,    
D.K.~Abhayasinghe$^\textrm{\scriptsize 93}$,    
S.H.~Abidi$^\textrm{\scriptsize 167}$,    
O.S.~AbouZeid$^\textrm{\scriptsize 40}$,    
N.L.~Abraham$^\textrm{\scriptsize 156}$
% ...
\bigskip
\\

% ...
$^{13}$Institute of Physics, Azerbaijan Academy of Sciences, Baku; Azerbaijan.\\
% ...
$^{40}$Niels Bohr Institute, University of Copenhagen, Copenhagen; Denmark.\\
% ...
$^{53}$II. Physikalisches Institut, Georg-August-Universit\"{a}t G\"ottingen, G\"ottingen; Germany.\\
% ...
$^{70}$$^{(a)}$INFN Sezione di Pavia;$^{(b)}$Dipartimento di Fisica, Universit\`a di Pavia, Pavia; Italy.\\
% ...
$^{93}$Department of Physics, Royal Holloway University of London, Egham; United Kingdom.\\
% ...
$^{101}$CPPM, Aix-Marseille Universit\'e, CNRS/IN2P3, Marseille; France.\\
$^{102}$Department of Physics, University of Massachusetts, Amherst MA; United States of America.\\
% ...
$^{128}$Homer L. Dodge Department of Physics and Astronomy, University of Oklahoma, Norman OK; United States of America.\\
% ...
$^{156}$Department of Physics and Astronomy, University of Sussex, Brighton; United Kingdom.\\
% ...
$^{167}$Department of Physics, University of Toronto, Toronto ON; Canada.\\
% ...
$^{a}$ Also at Borough of Manhattan Community College, City University of New York, New York NY; United States of America.\\
$^{b}$ Also at CERN, Geneva; Switzerland.\\
% ...
$^{*}$ Deceased

\end{flushleft}

 
\end{document}